\documentclass[apjl]{emulateapj}

\usepackage{comment}
\usepackage{ifthen}
\usepackage{amsmath}
\usepackage{amsfonts}
\usepackage{amssymb}
\usepackage{subfigure}
\usepackage{epsfig}
\usepackage{wasysym}

\newcommand{\luna}{{\tt LUNA}}
\newcommand{\multi}{{\sc MultiNest}}

\newcommand{\cofiam}{{\tt CoFiAM}}
\newcommand{\kic}{KIC 10593626}

\newcommand{\koi}{KOI-87}

\newcommand{\kepler}{Kepler-22}
\newcommand{\keplerb}{Kepler-22b}

\newboolean{emulateapj}
\setboolean{emulateapj}{true}

\newboolean{astroph}
\setboolean{astroph}{true}

\shortauthors{The HEK Project}
\shorttitle{The First Search for a Habitable Exomoon}
\ifthenelse{\boolean{emulateapj}}{
    \newcommand{\titledag}{$\dagger$}
}{
    \newcommand{\titledag}{\dagger}
}

\begin{document}

\title {The Hunt for Exomoons with Kepler (HEK):\\
 III. The First Search for an Exomoon around a Habitable-Zone Planet \altaffilmark{\titledag}}

\author{
	{\bf 
D.~M.~Kipping\altaffilmark{1,2},
D.~Forgan\altaffilmark{3},
J.~Hartman\altaffilmark{4},\\
D.~Nesvorn\'{y}\altaffilmark{5}, 
G.~\'A.~Bakos\altaffilmark{4}, 
A.~Schmitt\altaffilmark{6},
L.~Buchhave\altaffilmark{7}}
}
\altaffiltext{1}{Harvard-Smithsonian Center for Astrophysics,
	Cambridge, MA~02138, USA; dkipping@cfa.harvard.edu}

\altaffiltext{2}{Carl Sagan Fellow}

\altaffiltext{3}{Scottish Universities Physics Alliance (SUPA), Institute for
Astronomy, University of Edinburgh, Blackford Hill, Edinburgh, EH9 3HJ, 
Scotland, UK}

\altaffiltext{4}{Department of Astrophysical Sciences, Princeton University, 
Princeton, NJ~05844, USA}

\altaffiltext{5}{Department of Space Studies, Southwest Research Institute, 
Boulder, CO~80302, USA}

\altaffiltext{6}{Citizen Science}

\altaffiltext{7}{Niels Bohr Institute, Copenhagen University, Denmark}

\altaffiltext{$\dagger$}{
Based on archival data of the \emph{Kepler} telescope. 
}


\begin{abstract}

\keplerb\ is the first transiting planet to have been detected in the 
habitable-zone of its host star. At 2.4\,$R_{\oplus}$, \keplerb\ is too large to 
be considered an Earth-analog, but should the planet host a moon large enough to 
maintain an atmosphere, then the \kepler\ system may yet possess a telluric
world. Aside from being within the habitable-zone, the target is attractive due
to the availability of previously measured precise radial velocities and low
intrinsic photometric noise, which has also enabled asteroseismology studies of
the star. For these reasons, \keplerb\ was selected as a target-of-opportunity 
by the ``Hunt for Exomoons with Kepler'' (HEK) project. In this work, we
conduct a photodynamical search for an exomoon around \keplerb\ leveraging
the transits, radial velocities and asteroseismology plus several new tools
developed by the HEK project to improve exomoon searches. We find no evidence
for an exomoon around the planet and exclude moons of mass 
$M_S>0.5$\,$M_{\oplus}$ to 95\% confidence. By signal injection and blind 
retrieval, we demonstrate that an Earth-like moon is easily 
detected for this planet even when the time-correlated noise of the data set is 
taken into account. We provide updated parameters for the planet \keplerb\ 
including a revised mass of $M_P<53$\,$M_{\oplus}$ to 95\% confidence and an 
eccentricity of $0.13_{-0.13}^{+0.36}$ by exploiting Single-body 
Asterodensity Profiling (SAP). Finally, we show that \keplerb\ has a $>95$\% 
probability of being within the empirical habitable-zone but a $<5$\% 
probability of being within the conservative habitable-zone.

\end{abstract}



\section{INTRODUCTION}
\label{sec:intro}

\keplerb\ is a recently validated extrasolar planet detected by the \emph{Kepler
Mission} via the transit technique (\citealt{borucki:2012}, B12 hereafter). 
Orbiting a star with 75\% of solar luminosity once every 290\,days, \keplerb\ 
receives an insolation just 10-15\% greater than that received by the Earth and 
thus was claimed to be the first transiting planet discovered to orbit within 
the so-called habitable-zone of its parent sun (B12). Although habitable-zone 
exoplanets have been detected by radial velocity surveys 
\citep{vogt:2010,GJ667C:2012}, \keplerb\ is a fascinating object thanks to the 
plethora of follow-up opportunities afforded by transiting planets 
\citep{winn:2010}. This excitement is evident in the preceding astronomical 
literature with speculation about the planet's habitability 
\citep{neubauer:2012}, a presently undetected exomoon's habitability 
\citep{hellerbarnes:2012} and considerable efforts to redefine the so-called 
habitable-zone (e.g. \citealt{kopparapu:2013,vladilo:2013,zsom:2013}).

Despite the excitement generated by the pioneering discovery of \keplerb\ by
B12, the planet cannot be considered ``Earth-like'' given its
radius of 2.4\,$R_{\oplus}$. Observationally, the composition of the planet
is essentially unconstrained since the upper limit on the planetary mass is 
quite broad at $M_P<82$\,$M_{\oplus}$ (2\,$\sigma$) (B12). The
most well-characterized planet with a radius similar to \keplerb\ would seem to
be GJ\,1214b \citep{charbonneau:2009} with a radius of 2.7\,$R_{\oplus}$, which 
has a low bulk density indicating that it cannot have a rock-dominated 
composition \citep{escude:2012,mah:2013}.

Although \keplerb\ may not be a rocky habitable Earth-like world, there is a
distinct possibility for the system to yet maintain such a world if \keplerb\
possesses a large moon. Recently, there has been considerable speculation
about the potential habitability of such a moon \citep{hellerbarnes:2012,
heller:2013} since it is thought that exomoons should be detectable around 
transiting planets using Kepler \citep{kipping:2009a,kipping:2009b,
kipping:2009}. Despite this speculation, we note that, to our knowledge, no-one 
has ever conducted a search for an exomoon in this system. Indeed, we note that 
there has never even been a search for an exomoon around a habitable-zone 
\emph{candidate} planet, let alone a validated one, in the present literature.

The ``Hunt for Exomoons with Kepler'' (HEK) project \citep{hek:2012} is 
presently the only systematic program we are aware of attempting to 
observationally identify extrasolar satellites. Searching for such signals is 
computationally challenging due to the highly multimodal, complex and 
high-dimensional parameter space one is faced with \citep{hek:2012}, the need 
for calculating the Bayesian evidence integral \citep{hek:2013} and the 
photodynamic forward modeling required to simulate exomoon signals 
\citep{luna:2011}. To provide some context, the light curve fits presented in 
this work required 49.7\,years of CPU time\footnote{Using AMD Opteron 6272 \&
6282 SE processors}. To date, eight Kepler Objects of Interest (KOIs) have been 
analyzed for evidence of extrasolar moons each with null detections leading to 
upper limits on the satellite-to-planet mass ratio of $(M_S/M_P)\lesssim1$-$4$\% 
\citep{nesvorny:2012,hek:2013}.

In this work, we present an analysis of the first habitable-zone planet by
HEK. This target was identified as a target-of-opportunity by the project
because it i) is in the habitable-zone ii) is a validated planet iii) has
radial velocity measurements iv) has a quiet, bright ($K_P=11.7$) host star 
v) has a host star with asteroseismology constraints (B12). In 
this work, we include several new modes to thoroughly explore the exomoon 
parameter space. These include retrograde (\S\ref{sub:retro}) and eccentric 
moon solutions (\S\ref{sub:eccentricity}), informative and uninformative limb 
darkening priors (\S\ref{sub:LDpriors}), applying Bayesian model averaging 
(\S\ref{sub:BMA}), high-resolution fitting (\S\ref{sub:hires}), leveraging the 
radial velocities (\S\ref{sub:RVs}) and asteroseismology constraints 
(\S\ref{sub:astero}) and injecting and retrieving synthetic moon signals 
(\S\ref{sec:injected}). We also take this opportunity to provide updated
constraints on the mass, radius, composition, habitability and orbit of
\keplerb\ using new \emph{Kepler} data (\S\ref{sec:planetonly}).

\section{DATA HANDLING}
\label{sec:data}

\subsection{Data Acquisition}
\label{sub:dataacquisition}

In the discovery paper of \keplerb\ (B12), three transits were
detected by the \emph{Kepler Mission} occurring in quarters 1, 4 and 7 (Q1, Q4 
and Q7). Since this time, an additional three transits should have been 
observed by \emph{Kepler}: a $4^{\mathrm{th}}$ transit in Q11, a 
$5^{\mathrm{th}}$ transit in Q14, and a $6^{\mathrm{th}}$ transit in Q17. The 
$4^{\mathrm{th}}$ transit was successfully recorded but unfortunately the 
$5^{\mathrm{th}}$ was not due to a data gap in the Q14 time series. The 
$6^{\mathrm{th}}$ transit is unlikely to have been recorded because it occurred 
during a safe mode event from May $1^{\mathrm{st}}$ to May $6^{\mathrm{th}}$ 
2013. Further, soon after this \emph{Kepler} lost functionality of a second 
reaction wheel on May $12^{\mathrm{th}}$, putting future science observations in 
doubt. Therefore, it is quite possible that the four transits of \keplerb\ 
analyzed in this paper will be the only transits ever observed by \emph{Kepler}. 
The first transit was observed in long-cadence (LC) mode only, whilst the latter 
three have short-cadence (SC) data.

We downloaded the reduced data from the Mikulski Archive for Space Telescopes 
(MAST). In this work, we always make use of the ``raw'' (labelled as 
``SAP\_FLUX'' in the header) data processed by the Data Analysis Working Group 
(DAWG) pipeline (see accompanying data release notes for details). The ``raw'' 
data has been processed using PA (Photometric Analysis), which includes cleaning 
of cosmic ray hits, Argabrightenings, removal of background flux, aperture 
photometry and computation of centroid positions. For the sake of brevity, we do 
not reproduce the details of the PA pipeline here, but direct those interested 
to \citet{gilliland:2010} and the data release handbooks.

\subsection{Detrending with \cofiam}
\label{sub:cofiam}

The \emph{Kepler} photometry contains several systematic effects which require
detrending before a precise transit light curve analysis can be conducted.
These effects can be instrumental (such as focus drift, pointing tweaks,
safe modes, etc) or astrophysical (such as flaring, rotational modulations, 
etc). In this work, we utilize the \cofiam\ (Cosine Filtering with 
Autocorrelation Minimization) algorithm described in \citet{hek:2013} for
detrending these various effects.

\cofiam\ can be thought of as a high-pass, low-cut periodic filter optimized to
undisturb all periodicities at or below the so-called ``protected timescale'', 
$\mathfrak{T}$. The algorithm builds upon initial applications of cosine
filtering by \citet{mazeh:2010} for CoRoT photometry and \citet{kipbak:2011a,
kipbak:2011b} for \emph{Kepler} photometry. \cofiam\ regresses the following
sum of harmonic functions to the time series:

\begin{align}
F_k(t_i) = a_0 + \sum_{k=1}^{N_{\mathrm{order}}} \Bigg[ x_k \sin\Big(\frac{2\pi t_i k}{2D} \Big) + y_k \cos\Big(\frac{2\pi t_i k}{2D} \Big) \Bigg],
\label{eqn:linearcosinefilter}
\end{align}

where $D$ is the total baseline of the data under analysis, $t_i$ are the time
stamps of the data, $x_k$ \& $y_k$ are model variables and $N_{\mathrm{order}}$ 
is the highest harmonic order. For any given $\mathfrak{T}$,
one may define the maximum number of distinct harmonic cosines functions 
($N_{\mathrm{order}}^{\mathrm{max}}$) to regress to the data without disturbing 
the timescale $\mathfrak{T}$ as:

\begin{align}
N_{\mathrm{order}}^{\mathrm{max}} &= \frac{2 D}{4 \mathfrak{T}}.
\label{eqn:Nordermax}
\end{align}

However, choosing any $N_{\mathrm{order}}<N_{\mathrm{order}}^{\mathrm{max}}$ 
will also protect the timescale $\mathfrak{T}$. Exploiting this fact, \cofiam\ 
explores every possible variation (in total there are 
$N_{\mathrm{order}}^{\mathrm{max}}-1$ variations). The detrended light curve 
after each \cofiam\ regression is trimmed to within twice the ``window'' 
timescale ($T_{\mathrm{window}}$) of the predicted transit time and then a final
linear slope is fitted through the data (excluding the transit itself) to serve
as a final normalization.

After every detrending, we compute the autocorrelation on a 30\,minute timescale 
using the Durbin-Watson metric:

\begin{align}
d &= \frac{ \sum_{i=2}^N (r_i - r_{i-1})^2 }{ \sum_{i=1}^N r_i^2 },
\label{eqn:durbin}
\end{align}

where $r_i$ are the residuals and $N$ is the number of data points.
The value of $d$ always lies between 0 and 4, with 2 representing an absence of
autocorrelation, $|d-2|>0$ implying otherwise. \cofiam\ therefore selects the
value of $N_{\mathrm{order}}$ which minimizes $|d-2|$ (hence ``autocorrelation
minimization'').

\cofiam\ is applied to each transit epoch individually and the inputted data is 
pre-screened for any sharp, discontinuous offsets (due to say a pointing tweak)
as well as a general outlier rejection phase using median filtering. 
We never attempt to stitch two quarters together since there is no reason to 
expect the detrending function to be smooth between rolls of the spacecraft 
(which occur every quarter). 

In this work, we chose $\mathfrak{T} = 3 T_{14}$ where $T_{14}$ is the 
first-to-fourth contact transit duration. \citet{waldmann:2012} showed that
$T_{14}$ is the lowest periodicity in the Fourier transform of a transit and so
we multiply this by three to provide a small cushion. The ``window'' timescale
is defined to be $T_{\mathrm{window}} = 1.2 [(T_{14}/2) + T_{\mathrm{Hill}} + 
T_{14}]$ (the 1.2 factor is again a cushioning factor) where $T_{\mathrm{Hill}}$
is the Hill sphere timescale. For this we define $T_{\mathrm{Hill}} = 
[M_P/(3M_*)]^{1/3} [P_{B*}/(2\pi)]$ where $P_{B*}$ is the planet's orbital 
period\footnote{We use the subscript ``B*'' to denote that this term technically
refers to the barycentre (B) of a planet-moon system orbiting a star (*). For
an isolated planet, the subscripts ``B*'' and ``P'' are equivalent.}, $M_P$ is 
the planet's mass and $M_*$ is the stellar mass. For all of these quantities we 
simply used the best-fit values quoted in B12 (for $M_P$ we 
used the 3\,$\sigma$ upper mass limit).

We stress that the detrending function is regressed independently of
the later transit fits. This is done because the form of the detrending 
function is varied and we typically try between 
$N_{\mathrm{order}}^{\mathrm{max}}\sim10$-$30$ different models per transit. To 
perform the detrending in conjunction with the transits would thus require
around $N_{\mathrm{transits}} {^\wedge} N_{\mathrm{order}}^{\mathrm{max}}$ 
unique models i.e. $\mathcal{O}[10^5]$ for \keplerb. Given the very costly 
compuational demands of even a single photodynamical fit, this would be 
unrealistic with current computational capabilities.

\begin{figure*}
\subfigure[Quarter 1
\label{fig:Q1raw}]
{\epsfig{file=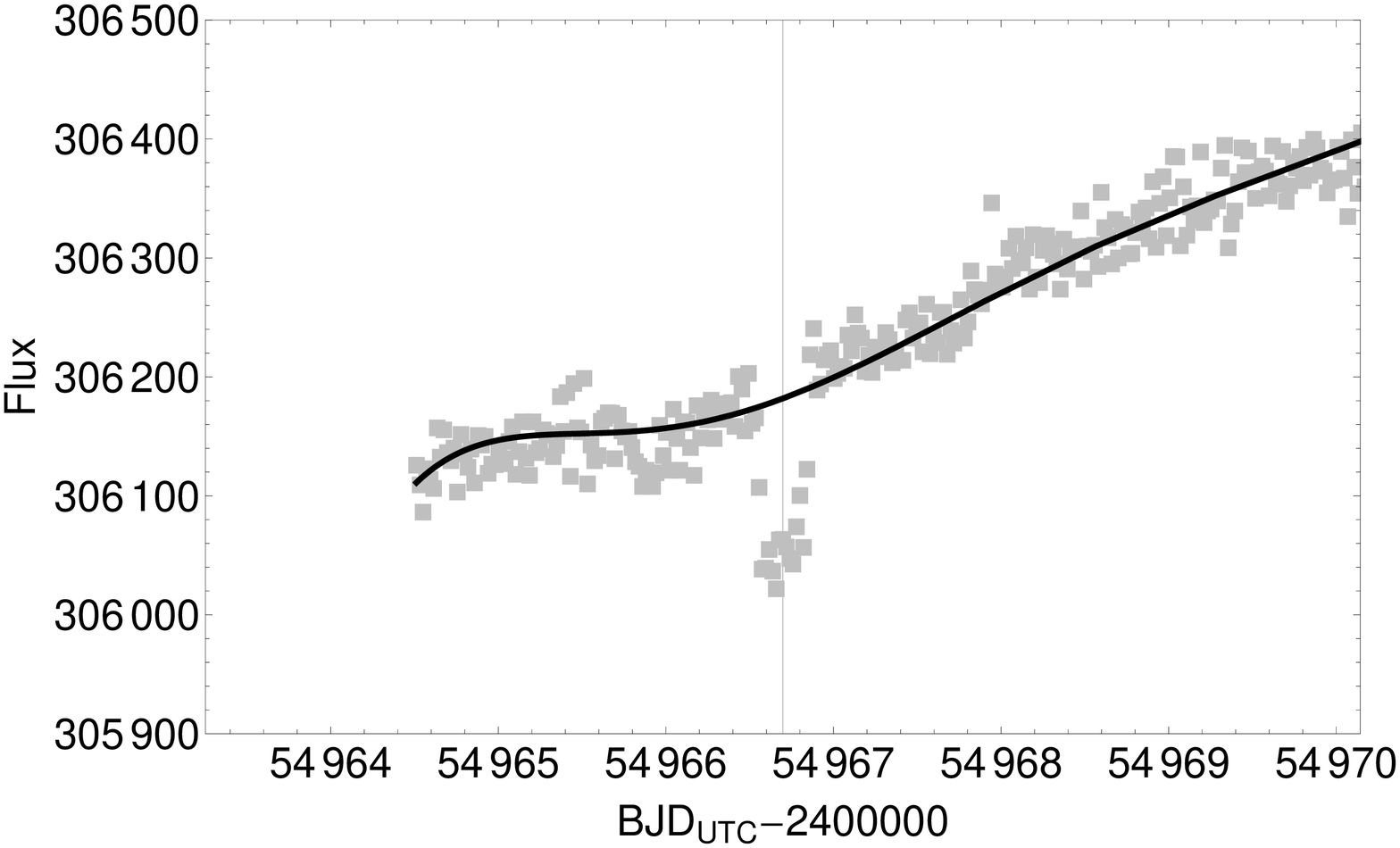,width=84mm}}
\subfigure[Quarter 4
\label{fig:Q4raw}]
{\epsfig{file=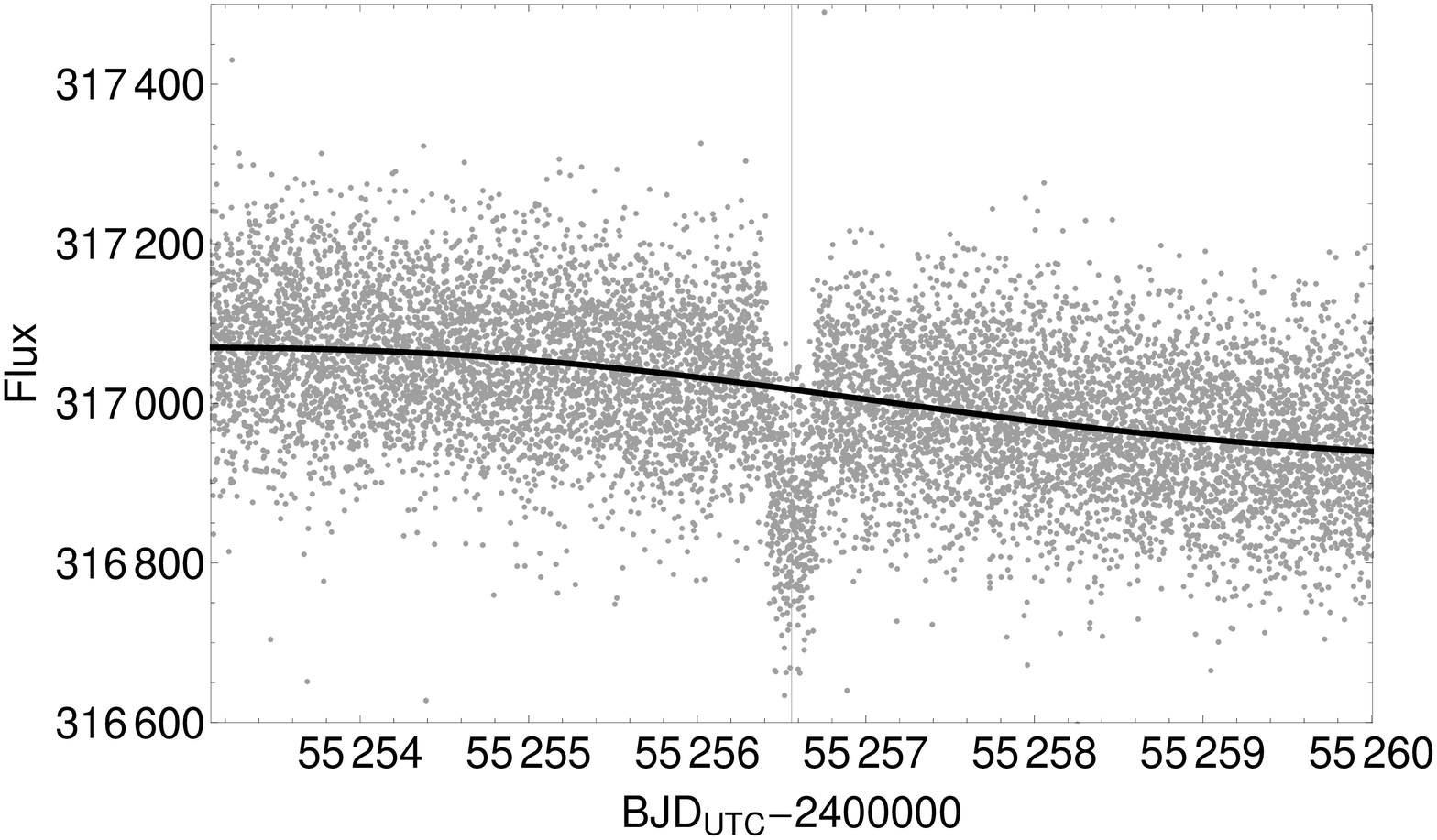,width=84mm}}\\
\subfigure[Quarter 7
\label{fig:Q7raw}]
{\epsfig{file=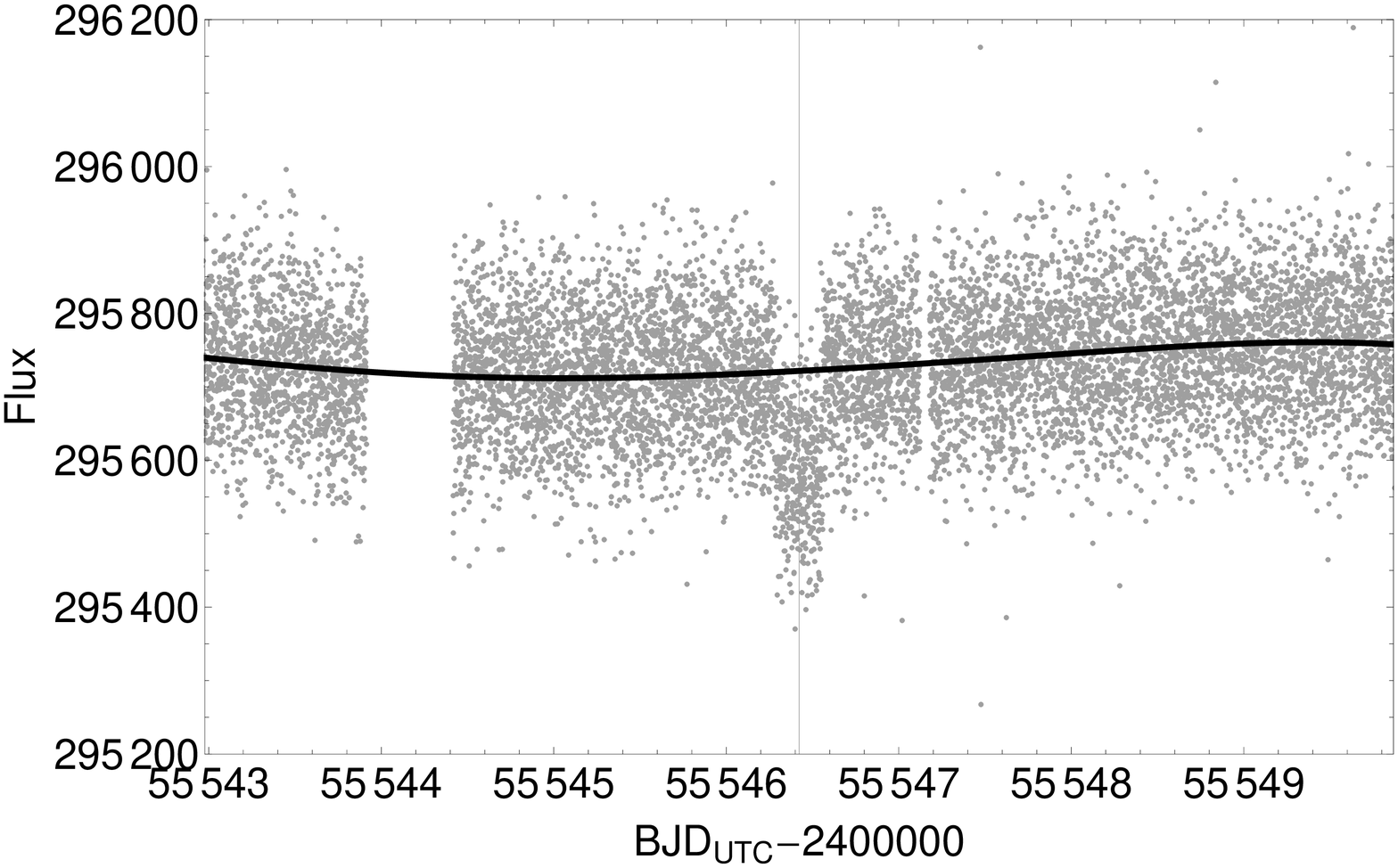,width=84mm}}
\subfigure[Quarter 11
\label{fig:Q11raw}]
{\epsfig{file=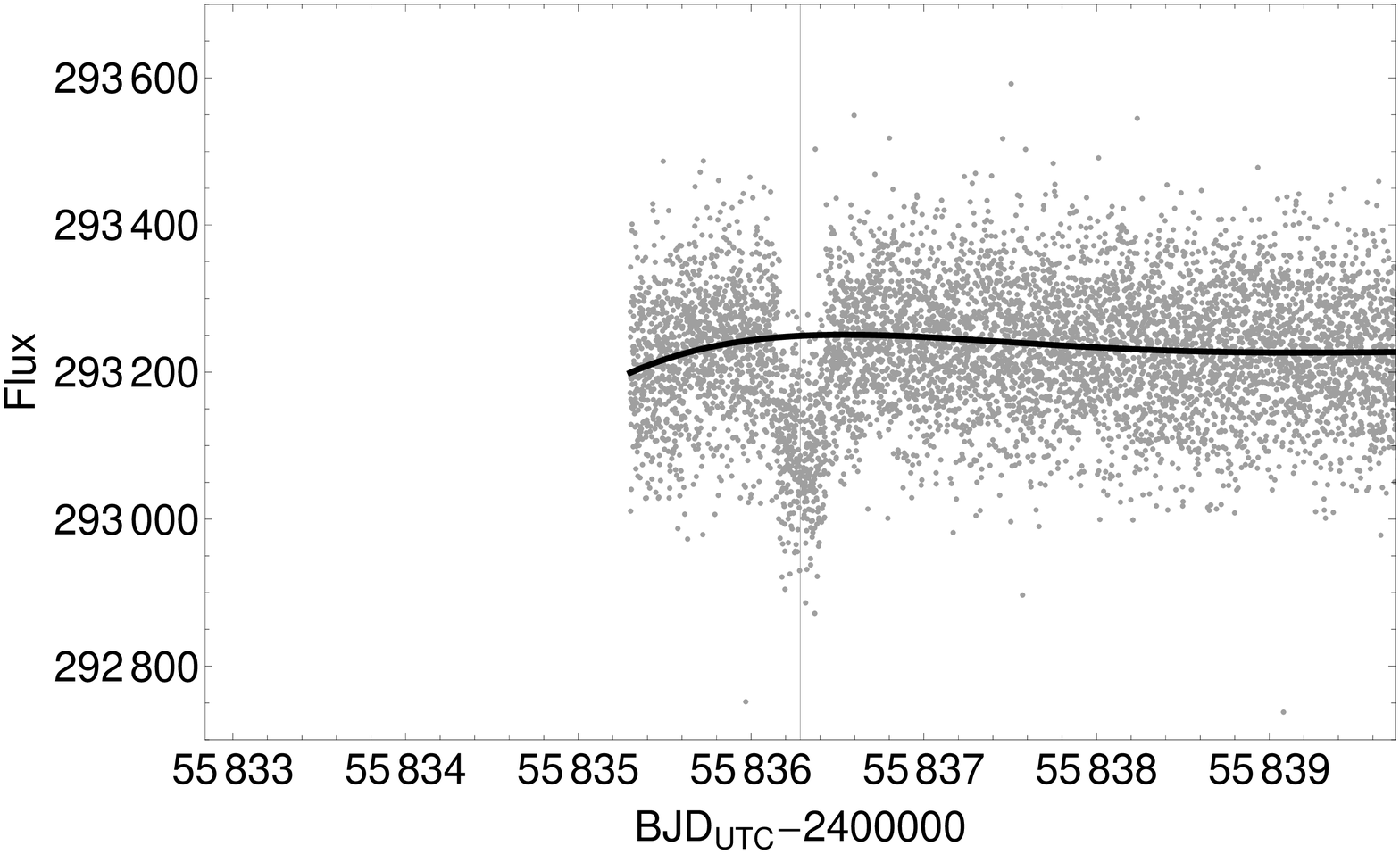,width=84mm}}\\
\caption{
``Raw'' (PA output) flux observed by \emph{Kepler} for \keplerb. Overlaid is our 
model for the long-term trend, computed by \cofiam. Long-cadence data 
(squares) only is available for Q1 but short-cadence data (dots) is available 
for the others. The location of the transits are marked with vertical 
gridlines.
\label{fig:QAraw}}
\end{figure*}

After detrending with \cofiam, we find $d_{Q1}=1.74069$, $d_{Q4}=1.97479$, 
$d_{Q7}=1.96775$ and $d_{Q11}=1.86203$ and the detrending functions are
plotted over the PA data in Figure~\ref{fig:QAraw}.

\section{MODELING}
\label{sec:modeling}

\subsection{Light Curve Fits}

We first fit and model the light curve as being due to a single planet
transiting the host star, to serve as a benchmark for the subsequent 
planet-with-moon fits. The planet-only light curve is modeled as was done in
\citet{hek:2013}, using the popular \citet{mandel:2002} algorithm. We assume a
quadratic limb darkening law for the host star, where the specific intensity
is described by $I(\mu) = 1 - u_1 (1-\mu) - u_2 (1-\mu)^2$ where 
$\mu=\sqrt{1-r^2}$, $r$ is the normalized radial coordinate on the stellar
disk and $u_1$ \& $u_2$ are the limb darkening coefficients.

Planet-with-moon models are generated using the photodynamic \luna\ algorithm
\citep{luna:2011}, which inherently accounts for the various timing effects
on the planet induced by a moon \citep{kipping:2009a,kipping:2009b}. We 
direct the reader to \citet{luna:2011} for details on how the photodynamical
light curves are computed.

Model light curves are fitted to the data using the multimodal nested
sampling algorithm \multi\ \citep{feroz:2009a,feroz:2009b}. By using this code,
we not only obtain parameter posteriors but also the Bayesian evidence of each
model attempted, thus allowing for later Bayesian model selection and model
averaged posteriors. Bayesian model selection is a crucial tool in detecting
exomoons due to the roughly doubling of free parameters combined with the
complex parameter space \citep{hek:2012}.

In general, the methods described above are the same as those implemented in
our previous HEK surveys (e.g. see \citealt{hek:2013}). Notably, the choice of
priors follows that of \citet{hek:2013} and are shown in Table~\ref{tab:priors}.
However, in studying \keplerb, we introduced several new changes to our 
methodology to improve both the accuracy of our derived results and the speed of 
the computations. These modifications are discussed in what follows.

\subsection{Priors for the Limb Darkening Coefficients}
\label{sub:LDpriors}

In this work, we use two different priors for the limb darkening coefficients:
i) informative priors ii) uninformative priors. The informative priors are
computed using a Monte Carlo forward modeling of stellar atmosphere models.
First, we draw a random normal variate from the effective temperature 
($T_{\mathrm{eff}}$) and stellar surface gravity ($\log g_*$) and then
compute the associated quadratic limb darkening coefficients using the 
\citet{kurucz:2006} stellar atmosphere model database convolved with the 
\emph{Kepler} bandpass (see \citet{kipbak:2011a} for more details\footnote{This
calculation is performed by a Fortran code written by I.~Ribas}). We repeat this
process until $10^4$ fair realizations of $u_1$ and $u_2$ have been computed.
For $T_{\mathrm{eff}}$ and $\log g_*$, we used the quoted values from 
B12 of $T_{\mathrm{eff}} = 5518\pm44$\,K and $\log g_* = 4.44 
\pm 0.06$ but doubled the uncertainties for both.

\begin{figure*}
\begin{center}
\includegraphics[width=16.0 cm]{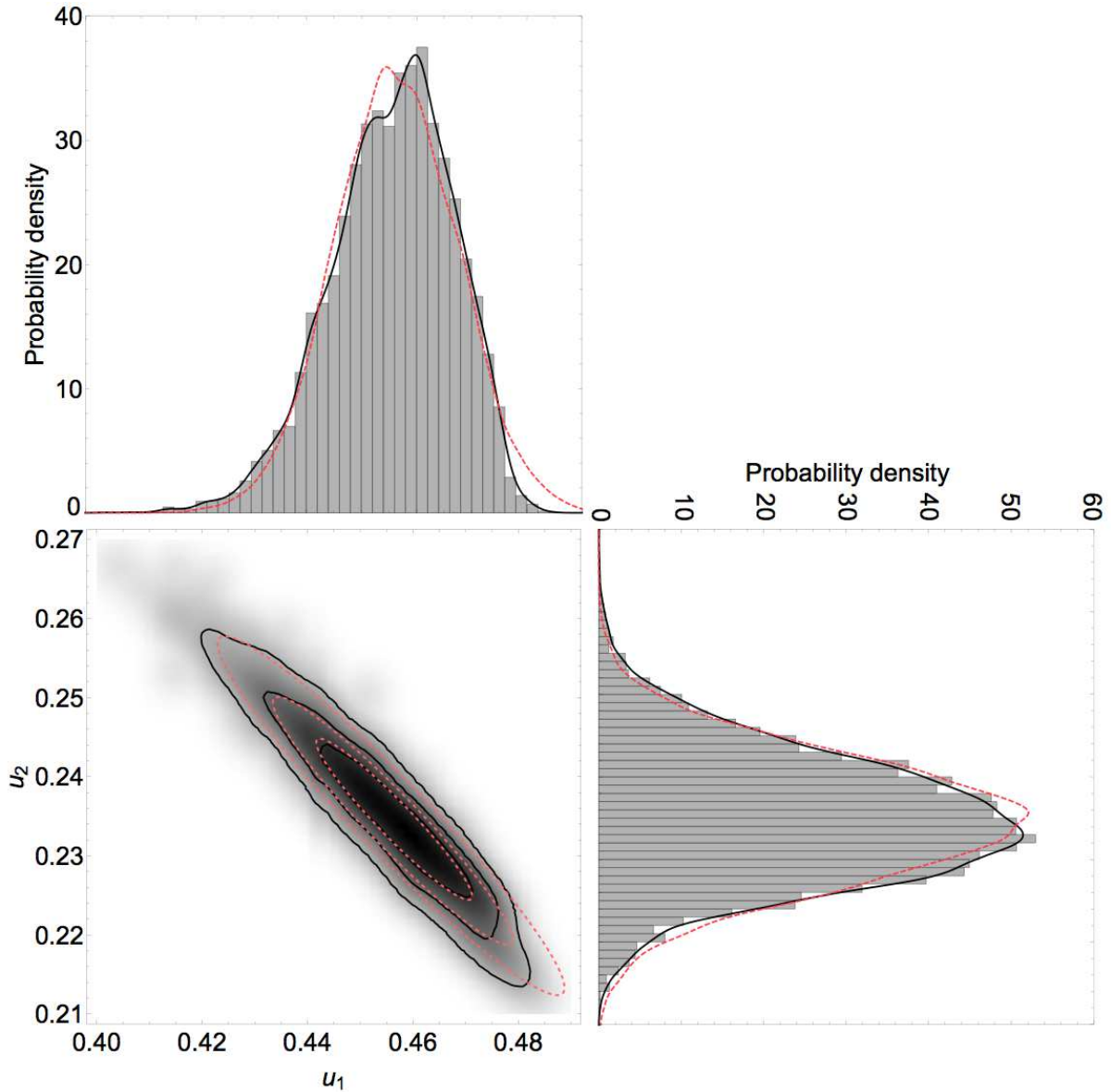}
\caption{\emph{Histograms show the prior probability density distribution of
the quadratic limb darkening coefficients $u_1$ and $u_2$, computed from
stellar atmosphere models. In black-solid we overlay the smoothed histogram of 
the same data. The red-dashed shows the analytic prior probability density
function predicted using our $w_1$-$w_2$ PCA transformed normal bivariate
prior probability density function (see \S\ref{sub:LDpriors} for details),
which serves as an informative prior in the transit fits. The bottom-left
panel shows the same as a joint-probability distribution, where
the contours represent the 1, 2 and 3\,$\sigma$ confidence limits.}} 
\label{fig:u1u2prior}
\end{center}
\end{figure*}

The $10^4$ realizations of $u_1$-$u_2$ form a joint prior probability for the
limb darkening coefficients. For computational expedience, it is desirable to 
characterize this joint prior probability with a simple analytic form, such as a 
bivariate Gaussian, rather than calling a stellar atmosphere model at every
realization in the light curve fits. However, a bivariate Gaussian is not 
ideal since our $u_1$-$u_2$ joint prior probability displays a strong 
covariance (correlation coefficient of -0.97). Despite this, a bivariate 
Gaussian can be applied by re-parameterizing the limb darkening coefficients. 
To this end, we perform a principal components analysis (PCA) on the joint prior 
probability of $u_1$-$u_2$ and re-parameterize the terms into the orthogonal 
components $w_1$ and $w_2$ following the suggestion of \citet{pal:2008}:

\begin{align}
w_1 &= u_1 \cos\varphi - u_2 \sin\varphi, \nonumber \\
w_2 &= u_2 \cos\varphi+ u_1 \sin\varphi,
\label{eqn:w1w2}
\end{align}

where we derived $\varphi = 34.56^{\circ}$ from the PCA. Using this 
transformation, we computed a joint prior probability for $w_1$-$w_2$ with
negligible covariance. The lack of covariance, combined with the fact the
resulting distributions are unimodal and approximately symmetric means that a 
bivariate Gaussian is now a reasonable analytic approximation. Regressing a 
Gaussian to each cumulative density function (CDF), we derive normal priors of 
$\mathrm{P}(w_1) \sim \mathcal{N}(0.242,0.013)$ and $\mathrm{P}(w_2) \sim 
\mathcal{N}(0.45215,0.00070)$. In Figure~\ref{fig:u1u2prior}, we compare
the PDF in $u_1$ and $u_2$ predicted by these priors relative to the actual
prior probability initially generated, where one can see the good agreement.

The second type of prior we try is an uninformative one. \citet{dirichlet:2013}
recently showed that uniform priors in $u_1$-$u_2$ can be drawn, without ever 
sampling unphysical limb darkening coefficients, by applying the ``triangular 
sampling'' technique. This is achieved by re-parameterizing the coefficients to 
$q_1$-$q_2$ and sampling uniformly between zero and unity in each, where

\begin{align}
q_1 &= (u_1 + u_2)^2,\\
q_2 &= \frac{u_1}{2(u_1 + u_2)}.
\end{align}

Employing triangular sampling is approximately twice as efficient than the 
uninformative priors used previously in \citet{hek:2013}. This is because
previously we used uniform priors in $0<u_1<2$ and $0<(u_1+u_2)<1$, which spends
exactly half of its time sampling from unphysical (and thus rejected) parameter
ranges \citep{dirichlet:2013}.

\subsection{Linear Minimization of Baseline Parameters}
\label{sub:OOT}

In principle, \cofiam\ should provide a light curve with an out-of-transit
baseline flux equal to exactly unity. In practice, even with \cofiam's final
stage linear normalization, it is prudent to regress a baseline normalization
factor for each transit epoch simultaneous to the transit parameters. One major
benefit of doing so is that the uncertainty of the baseline normalization is
propagated into the uncertainty of the other transit parameters. In the past,
we have treated the normalization factor as a free parameter in \multi\ like
any other term. For planets with many transits epochs, this is problematic
since the regression now includes a very large number of free parameters
leading to laborious CPU times. The fact that a relatively minor nuisance
parameter causes this dramatic slow-down makes the situation somewhat 
impractical.

In this work, we introduce a new refinement to the HEK fitting strategy which
simply treats the baseline parameters ($\mathrm{OOT}_j$) as a set of nuisance 
parameters. Rather than marginalizing over these terms, every realization of the
transit model simply optimizes the baseline parameters to the data. This is
easily achieved using a weighted linear minimization which is both extremely
quick and guaranteed to always find the global minimum. In principle, one
should wish to marginalize over nuisance parameters rather than simply
maximize the likelihood, but in practice we find this procedure has little
effect on the other transit parameters and leads to a significant speed-up.

%
%

\subsection{Eccentricity Caps}
\label{sub:eccentricity}

Unlike previous surveys, we decided to investigate the potential for eccentric
orbits for both the planet and the moon due to the high-priority nature of
\keplerb. Eccentric fits always require dramatically more time to explore
due to i) the introduction of two new free parameters ii) the requirement to
solve Kepler's transcendental equation $\sim$trillions of times during the
course of a single fit. Eccentric orbits can also produce unphysical
scenarios, unless limits are enforced on the allowed range in $e_{B*}$
(eccentricity of the planet-satellite barycentre around the host star) and
$e_{SB}$ (eccentricity of the satellite around the planet-satellite barycentre).

First of all, parameter realizations reproducing collisions should be avoided.
We therefore enforce that $a_{SP} (1-e_{SB}) > 2 R_P$, i.e. the periastron
separation between the planet and the satellite exceeds two planetary radii
(we use two planetary radii since the satellite's radius is $R_S<R_P$).
We use two planetary radii because the satellite's radius can be up to one
planetary radius in our fits. Similarly, for the planetary orbits, we enforce
$a_{B*} (1-e_{B*}) > 2 R_*$.

Additionally, \citet{domingos:2006} showed that the maximum stable 
planet-satellite separation is attenuated for eccentric orbits. 
\citet{domingos:2006} showed that all satellites must conform to:

\begin{align}
f < 0.9309 (1 - 1.0764 e_{B*} - 0.9812 e_{SB}),
\end{align}

where $f$ is equal to the semi-major axis of the satellite
relative to the planet in units of the Hill radius. Under the assumption that
$M_S \ll M_P \ll M_*$, \citet{kipping:2009a} showed that $f^3 = 
(3P_{SB}^2/P_{B*}^2)$, where $P_{SB}$ is the period of the satellite around the
planet-satellite barycentre. This allows us to enforce the \citet{domingos:2006} 
condition as:

\begin{align}
(3 P_{SB}^2/P_{B*}^2)^{1/3} < 0.9309 (1 - 1.0764 e_{B*} - 0.9812 e_{SB})
\end{align}

These constraints are optimistic in the sense that the orbits are assumed to be
coplanar and the moon is simply a test particle. More realistic solutions
can be found by exploring the three-body solution with non-zero masses and
three-dimensional geometry (e.g. see \citealt{donnison:2010}), but we reserve
these more detailed dynamical investigations for the final posterior analyses.

We also enforce the same limits on the planet and satellite densities used in
\citet{hek:2013}. These various limits are all imposed by simply rejecting any
trials reproducing such behavior. Due to the computationally much higher cost
of exploring eccentric solutions, we only explored eccentric-planet or
eccentric-moon solutions but not both simultaneously, as such fits were found to
be prohibitively CPU expensive. Since the Hill stability of a putative moon
decreases strongly with increasing eccentricity for both terms, we do not
envisage it likely that this significantly hinders our detection ability.

\subsection{Retrograde Orbits}
\label{sub:retro}

In \citet{hek:2013} we did not explore retrograde orbits for the exomoon,
which \luna\ defines as being when $180^{\circ}<i_{SB}<360^{\circ}$\footnote{
This definition is selected so that a coplanar, prograde moon transits the
planet with an inclination of $90^{\circ}$, following the definition used for
a planet transiting a star.}. In this work, we present a method to explore such 
orbits whilst still maintaing an isotropic prior in $i_{SB}$. One difficultly in 
doing this is that an isotropic prior is uniform in $\cos i_{SB}$ but this 
trigonometric function is not uniquely defined over the interval 
$0^{\circ}<i_{SB}<360^{\circ}$. For example, $\cos i_{SB}=0$ may refer to 
$i_{SB}=90^{\circ}$ or $i_{SB}=270^{\circ}$.

Our resolution to this is to introduce an auxiliary cosine term which we dub
$\cos'i_{SB}$ and is defined over the interval $-1<\cos'i_{SB}<+3$. The inverse
of this function is given by:

\begin{equation*}
i_{SB} =
\begin{cases}
\cos^{-1}[\cos'i_{SB}]  & \text{if } -1 < \cos'i_{SB} \leq +1 ,\\
\cos^{-1}[\cos'i_{SB}-2]+\pi & \text{if } +1 < \cos'i_{SB} \leq +3.
\end{cases}
\label{eqn:inclination}
\end{equation*}

By enforcing a uniform prior in $-1<\cos'i_{SB}<+3$ and inverting the above, we 
reproduce an isotropic prior in inclination and still explore the full range of 
inclinations (i.e. both prograde and retrograde orbits).

\subsection{Asteroseismology Stellar Density Prior}
\label{sub:astero}

Unlike the previous KOIs analyzed in survey I \citep{hek:2013}, \kepler\ has
asteroseismology constraints available. Consequently, there exists a strong
empirical prior on $\rho_*$, the mean stellar density of 
$\rho_* = 1.458\pm0.030$\,g\,cm$^{-3}$ (B12). In all fits, we
treat $\rho_*$ as a normal prior following this measurement.

Usually, the light curve fitted eccentricity and stellar density are highly 
degenerate but the presence of a strong prior on the latter breaks this
degeneracy. This trick was first noted in \citet{map:2012} and was dubbed
``Single-body Asterodensity Profiling'' by the authors, or SAP. We are
therefore able to provide SAP-derived eccentricity constraints in the light
curve fits provided in this work.

We also note that B12 are able to combine their precise
$\rho_*$ measurement with stellar evolution isochrone models to provide
precise estimates of the stellar radius and mass ($R_*$ and $M_*$).
Although these are not model parameters in any our fits, we can use them
later to derive physical parameters.

\subsection{Radial Velocities (RVs)}
\label{sub:RVs}

As part of the campaign to validate the planetary nature of \keplerb, 
B12 obtained sixteen high resolution spectra of the target
star between 17 August 2010 and 25 August 2011 using the HIRES spectrometer
on the Keck I 10\,m telescope, which we utilize in this work. The spectra
yield $\sim1.4$\,m\,s$^{-1}$ precision radial velocities and show no clear 
detection of \keplerb. The null-detection is consistent with a low-mass planet, 
as expected for the radius of \keplerb. By leveraging these radial velocities,
we can constrain various parts of the parameter space explored by \multi, such
as excluding massive planets or moderate-mass planets on highly eccentric 
orbits.

Including the RVs requires only two extra free parameters to our model, $K$
and $s$ (the radial velocity jitter), since all of the other orbital parameters
are included in our fits anyway. The jitter term behaves like an additional
error which is simply added in quadrature to the reported RV uncertainties.
Unlike B12, we do not choose a fixed $s$ value of 
$3$\,m\,s$^{-1}$ but rather we fit the parameter using a modified Jeffrey's 
prior, as advocated by \citet{balan:2009}. Following \citet{balan:2009}, we 
choose the inflection point of the modified Jeffrey's prior to be equal to the 
median RV uncertainty and set the maximum to be equal to twice the range of the 
reported RVs. The same prior is also used for $K$.

The radial velocities can be used to derive the planetary mass, $M_P$, given a 
known stellar mass, $M_*$. To determine $M_P$, we solve the cubic equation of 
the well-known mass-function using Equation 3.18 of \citet{thesis:2011}. Whilst
some authors, such as B12, have advocated exploring negative
$K$ solutions to avoid positive-biases in fitting joint RV + transit data,
this is actually unnecessary for \multi\ as the nested sampling technique does
not suffer boundary condition biases, unlike Markov chains.

\subsection{Negative Radii Moons}
\label{sub:negativemoons}

A new feature we implement in this work for the HEK project is the exploration
of negative $(R_S/R_P)$ solutions i.e. negative radii moons. Negative radii 
moons are of course not physically plausible and our implementation simply 
involves flipping the transit signal component caused by the moon in such cases. 
We exploit this trick as a vetting test such that solutions favoring a negative
radius moon can be easily dismissed. This replaces the previous test of
\citet{hek:2013} of fitting for zero-radius moons and comparing the Bayesian
evidence. We find that negative-radius moon exploration is more efficient
computationally since we do not require another additional moon fit purely
for vetting purposes. We select a uniform prior on the radius ratio term of
$-1<(R_S/R_P)<+1$.

\begin{table}
\caption{\emph{Planet-moon parameters used in light curve fits and their 
associated priors. 
$\mathcal{U}\{a,b\}$ is a uniform prior between $a$ and $b$.
$\mathcal{N}\{a,b\}$ is a Gaussian prior with a mean of $a$ and standard 
deviation $b$.
$\mathcal{J}'\{a,b\}$ is a modified Jeffrey's prior with a maximum at $b$ and
an inflection point at $a$.
$^{*}$ represents uninformative limb darkening coefficient priors; informative
priors are discussed in \S\ref{sub:LDpriors}.
}} 
\centering 
\begin{tabular}{c c} 
\hline
Parameter & Prior \\ [0.5ex] 
\hline
\emph{Planet Parameters} & \\
\hline
$(R_P/R_*)$ & $\mathcal{U}\{0,1\}$ \\
$\rho_{*}^{\mathrm{circ}}$\,\,[$\mathrm{kg}$\,$\mathrm{m}^{-3}$] & $\mathcal{N}\{[1.458,0.030\}$ \\
$b_{B*}$ & $\mathcal{U}\{0,2\}$ \\
$P_{B*}$\,\,$[\mathrm{days}]$ & $\mathcal{U}\{288.8623,290.8623\}$ \\
$\tau_{B*}$\,\,[BJD$_{\mathrm{UTC}}$] & $\mathcal{U}\{2455545.4228,2455547.4228\}$ \\
$e_{B*}$ & $\mathcal{U}\{0,1\}$ \\
$\omega_{B*}$\,[rads] & $\mathcal{U}\{0,2\pi\}$ \\
$K$\,[m\,s$^{-1}$] & $\mathcal{J}'\{1.55,25.54\}$ \\
$s$\,[m\,s$^{-1}$] & $\mathcal{J}'\{1.55,25.54\}$ \\
$q_1^{*}$ & $\mathcal{U}\{0,1\}$ \\
$q_2^{*}$ & $\mathcal{U}\{0,1\}$ \\
\hline
\emph{Moon Parameters} & \\
\hline
$(R_S/R_P)$ & $\mathcal{U}\{-1,1\}$ \\
$(M_S/M_P)$ & $\mathcal{U}\{0,1\}$ \\
$(a_{SP}/R_P)$ & $\mathcal{U}\{2,345.9\}$ \\
$\cos'i_{SB}$\,\,[rads] & $\mathcal{U}\{-1,+3\}$ \\
$\Omega_{SB}$\,\,[rads] & $\mathcal{U}\{-\pi,+\pi\}$ \\
$P_{SB}$\,\,$[\mathrm{days}]$ & $\mathcal{U}\{0.052,167.4\}$ \\
$\phi_{SB}$\,\,[rads] & $\mathcal{U}\{0,2\pi\}$ \\
$e_{SB}$ & $\mathcal{U}\{0,1\}$ \\
$\omega_{SB}$\,[rads] & $\mathcal{U}\{0,2\pi\}$ \\ [1ex]
\hline\hline 
\end{tabular}
\label{tab:priors} 
\end{table}

\subsection{High-Resolution Fitting}
\label{sub:hires}

Due to the high priority nature of \keplerb, we implemented our fits
in a higher-than-usual resolution mode. There are two ways in which this is
implemented. Firstly, the integration time of the long-cadence data requires
correcting for using resampling \citep{binning:2010}. Typically, using a
resampling resolution of $N_{\mathrm{resam}} = 5$ is sufficient for most
\emph{Kepler} targets \citep{binning:2010,kipbak:2011a}. However, here we 
resample to a full short-cadence resolution using $N_{\mathrm{resam}} = 30$.

Additionally, the number of live points in \multi\ constrains how thoroughly the
code explores the parameter volume. \citet{feroz:2009b} recommend using 4000 
live points but we decided to double this to 8000 live points in what follows to 
ensure a thorough search for minima.

\subsection{Bayesian Model Averaging (BMA)}
\label{sub:BMA}

Altogether, ten different light curve models are regressed to the photometry
of \keplerb\ with varying underlying assumptions, as described in 
Table~\ref{tab:models}. Due to the considerable number of feasible models which 
we attempt to explain the data with, Bayesian model averaging (BMA) is a 
particularly powerful way of including our ignorance as to which of the 
attempted models is the correct one. In general, standard statistical practice 
ignores model uncertainty- an observer selects a model from some class of models 
and then proceeds under the assumption that the selected model generated the 
observations. This procedure ignores the model uncertainty and leads to 
over-confident inferences of the parameter posteriors. BMA provides a coherent 
approach for including our uncertainty in the models themselves.

\begin{table*}
\caption{\emph{
Description of the ten different models used in this work and three different
Bayesian model averages.
}} 
\centering 
\begin{tabular}{l l} 
\hline
Model & Description \\ [0.5ex] 
\hline
$\mathcal{P}_{\mathrm{LD-prior}}$	& Planet-only with $e_{B*}=0$ and informative LD prior \\
$\mathcal{P}_{\mathrm{LD-prior},e_{B*}}$& Planet-only with free $e_{B*}$ and uninformative LD prior \\
$\mathcal{P}_{\mathrm{LD-free}}$	& Planet-only with $e_{B*}=0$ and uninformative LD prior \\
$\mathcal{P}_{\mathrm{LD-free},e_{B*}}$	& Planet-only with free $e_{B*}$ and uninformative LD prior \\
$\mathcal{S}_{\mathrm{LD-prior}}$	& Planet-with-moon with $e_{B*}=e_{SB}=0$ and informative LD prior \\
$\mathcal{S}_{\mathrm{LD-prior},e_{B*}}$& Planet-with-moon with free $e_{B*}$, $e_{SB}=0$ and informative LD prior \\
$\mathcal{S}_{\mathrm{LD-prior},e_{SB}}$& Planet-with-moon with $e_{B*}=0$, free $e_{SB}=0$ and informative LD prior \\
$\mathcal{S}_{\mathrm{LD-free}}$	& Planet-with-moon with $e_{B*}=e_{SB}=0$ and uninformative LD prior \\
$\mathcal{S}_{\mathrm{LD-free},e_{B*}}$	& Planet-with-moon with free $e_{B*}$, $e_{SB}=0$ and uninformative LD prior \\
$\mathcal{S}_{\mathrm{LD-free},e_{SB}}$	&  Planet-with-moon with $e_{B*}=0$, free $e_{SB}=0$ and uninformative LD prior \\
\hline
$<\mathcal{P}_k>$		& Bayesian model average of all $\mathcal{P}_k$ models \\
$<\mathcal{P}_k,\mathcal{S}_k>$ for $e_{SB}=0$	& Bayesian model average of all models with $e_{SB}=0$ \\
$<\mathcal{P}_k,\mathcal{S}_k>$	& Bayesian model average of all models \\ [1ex]
\hline\hline 
\end{tabular}
\label{tab:models} 
\end{table*}

To our knowledge, BMA has not been previously applied in exoplanet studies but 
we here introduce the first application. A brief review of BMA and its 
applications to astrophysics is provided by \citet{parkinson:2013}. One may 
write the model-averaged posterior distribution for a given parameter $\Theta$
($\mathrm{P}(\Theta|\mathcal{D})$) as a function of the weighted sum of the 
posteriors from each individual model $\mathcal{M}_k$ 
($\mathrm{P}(\Theta|\mathcal{D},\mathcal{M}_k)$):

\begin{align}
\mathrm{P}(\bar{\Theta}|\mathcal{D}) = \frac{\sum_k 
\mathrm{P}(\Theta|\mathcal{D},\mathcal{M}_k) \mathrm{P}(\mathcal{M}_k|\mathcal{D})}{
\sum_k \mathrm{P}(\mathcal{M}_k|\mathcal{D}) },
\label{eqn:BAM1}
\end{align}

where $\mathcal{D}$ represents the data. The model weightings can be easily
defined using Bayes' theorem:

\begin{align}
\mathrm{P}(\mathcal{M}_k|\mathcal{D}) &= \frac{\mathrm{P}(\mathcal{D}|\mathcal{M}_k) \mathrm{P}(\mathcal{M}_k)}{\mathrm{P}(\mathcal{D})}.
\end{align}

The term $\mathrm{P}(\mathcal{D}|\mathcal{M}_k)$ is also known as the Bayesian
evidence ($\mathcal{Z}$) and is directly computed by \multi\ for each model 
attempted. The prior probability of each model $\mathcal{M}_k$ is assumed to 
be equal in what follows, and thus has no impact on the above expression. 
Likewise, $\mathrm{P}(\mathcal{D})$ is a normalization term which cancels out in 
Equation~\ref{eqn:BAM1}. Thus, we are left with:

\begin{align}
\mathrm{P}(\Theta|\mathcal{D}) = \frac{\sum_k 
\mathrm{P}(\Theta|\mathcal{D},\mathcal{M}_k) \mathcal{Z}_k}{\sum_k \mathcal{Z}_k}.
\label{eqn:BAM2}
\end{align}

\section{PLANET-ONLY RESULTS}
\label{sec:planetonly}

\subsection{Model Comparison}

For the planet-only models, denoted by $\mathcal{P}_i$, we tried four models
in total by exploring informative versus uninformative (free) priors on the
limb darkening coefficients and circular versus eccentric priors on the orbit.
The Bayesian evidences of the four models (see Table~\ref{tab:evidence}) are
broadly similar with the biggest $\Delta\log\mathcal{Z}$ occurring between
$\mathcal{P}_{\mathrm{LD-free},e_{B*}}$ and $\mathcal{P}_{\mathrm{LD-free}}$
where the former is favored at 1.2\,$\sigma$. This marginal and insignificant
difference suggests that the light curve contains insufficient information
to distinguish between these scenarios. We consider that the null-hypothesis
of a circular orbit is not over-turned by these results and we show the
maximum a-posteriori light curve fit from model 
$\mathcal{P}_{\mathrm{LD-free}}$ in Figure~\ref{fig:bestfits_es}.

The close proximity in evidence between the four planet-only models also
highlights the benefits of Bayesian model averaging (BMA), as discussed
earlier in \S\ref{sub:BMA}. In Table~\ref{tab:params}, columns two shows
the marginalzied parameters from model $\mathcal{P}_{\mathrm{LD-free},e_{B*}}$
alone and column three shows the effect of applying BMA over the four attempted 
planet-only models ($<\mathcal{P}_k>$).

\subsection{Transit Timing and Duration Variations (TTVs \& TDVs)}

In addition to the four planet-only fits discussed in the previous subsection,
we also tried three additional models designed to investigate the possibility
of dynamical timing variations. These fits seek to determine the transit times
and durations purely from the photometry and so we do not include the radial
velocity data in this fit, which theoretically adds some extra information
on the period and epoch of the transits. Since the input data is different,
we require a new null hypothesis fit for which we can compare subsequent
dynamical fits. To this end, we first tried a simple static transit model,
$\mathcal{P}'_{\mathrm{static}}$, where the dash is used to denote that this is
a different family of model classes to those tried earlier (and not directly
comparable).

$\mathcal{P}'_{\mathrm{static}}$ is identical to the 
$\mathcal{P}_{\mathrm{LD-prior}}$ model fit except the radial velocity data
is not included and so $K$ and $s$ are not free parameters in the fit. Next,
we try model $\mathcal{P}'_{\mathrm{TTV}}$, which allows each transit
epoch to have a unique transit time parameter. The period of such a fit is
unconstrained but does affect the estimate of $a_{B*}/R_*$ since this term is a
function of $P_{B*}$ and $\rho_*$. Therefore, we apply a Gaussian prior on this
term using the posterior distribution of $P_{B*}$ from model 
$\mathcal{P}'_{\mathrm{static}}$. Finally, we try model 
$\mathcal{P}'_{\mathrm{TTV+TDV}}$ in which each transit has entirely unique
transit parameters. 

\begin{figure*}
\begin{center}
\includegraphics[width=18.0 cm]{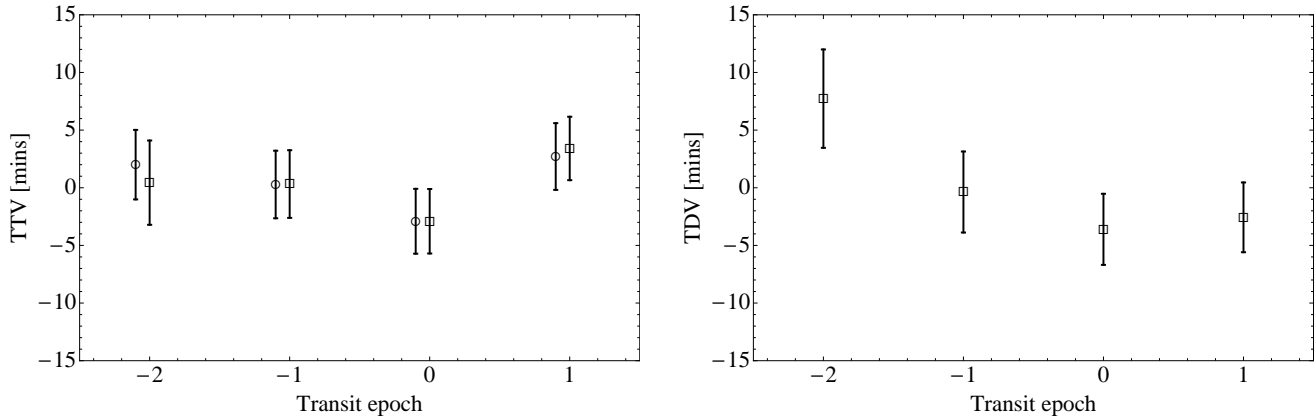}
\caption{\emph{Left panel: TTVs measured for \keplerb, relative to the maximum
a-posteriori ephemeris derived by model $\mathcal{P}'_{\mathrm{static}}$.
Squares represent the results determined by fitting each transit individually.
Circles (slightly offset in transit epoch for clarity) represent the results
determined by fitting global transit parameters but unique transit times. The
agreement in both accuracy and precision is excellent. Right panel: TDVs 
determined by the individual transit fitting model (it is not possible to 
provide TDVs when assuming global transit shape parameters). Durations defined 
using the $\tilde{T}$ definition and TDVs computed relative to the maximum 
a-posteriori duration from model $\mathcal{P}'_{\mathrm{static}}$.
}} 
\label{fig:TTVs}
\end{center}
\end{figure*}

After executing the fits in \multi, we find that a static model is strongly
favored over the competing hypothesis (see Table~\ref{tab:ttvevidence}). This
is supported by a simple analysis of the derived TTVs and TDVs, as shown in
Figure~\ref{fig:TTVs}, which displays a lack of any significant deviations
in either metric. We provide our derived transit times and durations in
Table~\ref{tab:TTVs}.

\begin{table*}
\caption{\emph{
Bayesian evidences of the three different models attempted to investigate
the possibility of dynamical variations in the transits of \keplerb. A static
orbit model is strongly favored over the competing hypotheses.
}} 
\centering 
\begin{tabular}{l c r c} 
\hline
Model & $\log\mathcal{Z}$ & Odds Ratio & $\sigma$ Confidence \\ [0.5ex] 
\hline
$\mathcal{P}'_{\mathrm{static}}$	& $176873.289 \pm 0.075$ & 1 & - \\
$\mathcal{P}'_{\mathrm{TTV}}$		& $176864.058 \pm 0.089$ & $9.8\times10^{-5}$ & $-3.90$ \\
$\mathcal{P}'_{\mathrm{TTV+TDV}}$	& $176847.848 \pm 0.120$ & $8.9\times10^{-12}$ & $-6.82$ \\ [1ex]
\hline\hline 
\end{tabular}
\label{tab:ttvevidence} 
\end{table*}

\begin{table*}
\caption{\emph{
Transit times and durations derived for \keplerb. The transit timing and 
duration variations, derived from these values, are shown in 
Figure~\ref{fig:TTVs}. These variations are computed relative the maximum
a-posteriori metrics from the static model $\mathcal{P}'_{\mathrm{static}}$;
specifically $\tau_0 = 2455546.4246$\,BJD$_{\mathrm{UTC}}$, 
$P_{B*} = 289.8645$\,days and $\tilde{T} = 419.4$\,mins.
}} 
\centering 
\begin{tabular}{c c c c} 
\hline
Epoch & $\tau_i$ [BJD$_{\mathrm{UTC}}$] from $\mathcal{P}'_{\mathrm{TTV}}$ & $\tau_i$ [BJD$_{\mathrm{UTC}}$] from $\mathcal{P}'_{\mathrm{TTV+TDV}}$ & $\tilde{T}$ [mins] from $\mathcal{P}'_{\mathrm{TTV+TDV}}$ \\ [0.5ex] 
\hline
-2	& $2454966.6969_{-0.0021}^{+0.0021}$ & $2454966.6958_{-0.0026}^{+0.0025}$ & $434.9_{-8.4}^{+8.6}$ \\
-1	& $2455256.5602_{-0.0020}^{+0.0020}$ & $2455256.5603_{-0.0021}^{+0.0020}$ & $418.7_{-7.2}^{+6.8}$ \\
0	& $2455546.4226_{-0.0020}^{+0.0019}$ & $2455546.4226_{-0.0019}^{+0.0020}$ & $412.2_{-6.2}^{+6.1}$ \\
+1	& $2455836.2910_{-0.0020}^{+0.0021}$ & $2455836.2915_{-0.0019}^{+0.0019}$ & $414.3_{-6.1}^{+6.0}$ \\ [1ex]
\hline\hline 
\end{tabular}
\label{tab:TTVs} 
\end{table*}

\subsection{Constraints on the Orbital Eccentricity}

Leveraging both single-body asterodensity profiling (SAP) \citep{map:2012}
and the radial velocities, we were able to allow for free eccentricity in
the light curve fits. For simplicity, we used a uniform prior in $e_{B*}$ and
$\omega_{B*}$ and discuss here the results from the model averaged planet-only 
posteriors (column heading $<\mathcal{P}_k>$ of Table~\ref{tab:params}). The 
$\Psi$ parameter (see Equation~6 of \citealt{map:2012}), which is a single 
measure of the eccentricity as determined using SAP \citep{map:2012}, is 
consistent with a circular orbit ($\Psi=1$) at $\Psi=1.4_{-0.6}^{+3.0}$. 
Further, the model averaged eccentricity is consistent with a circular orbit at 
$e_{B*}=0.13_{-0.13}^{+0.36}$ yielding a 95\% confidence upper limit of 
$e_{B*}<0.71$.

\subsection{Constraints on the Mass and Surface Gravity}

\begin{figure}
\begin{center}
\includegraphics[width=8.4 cm]{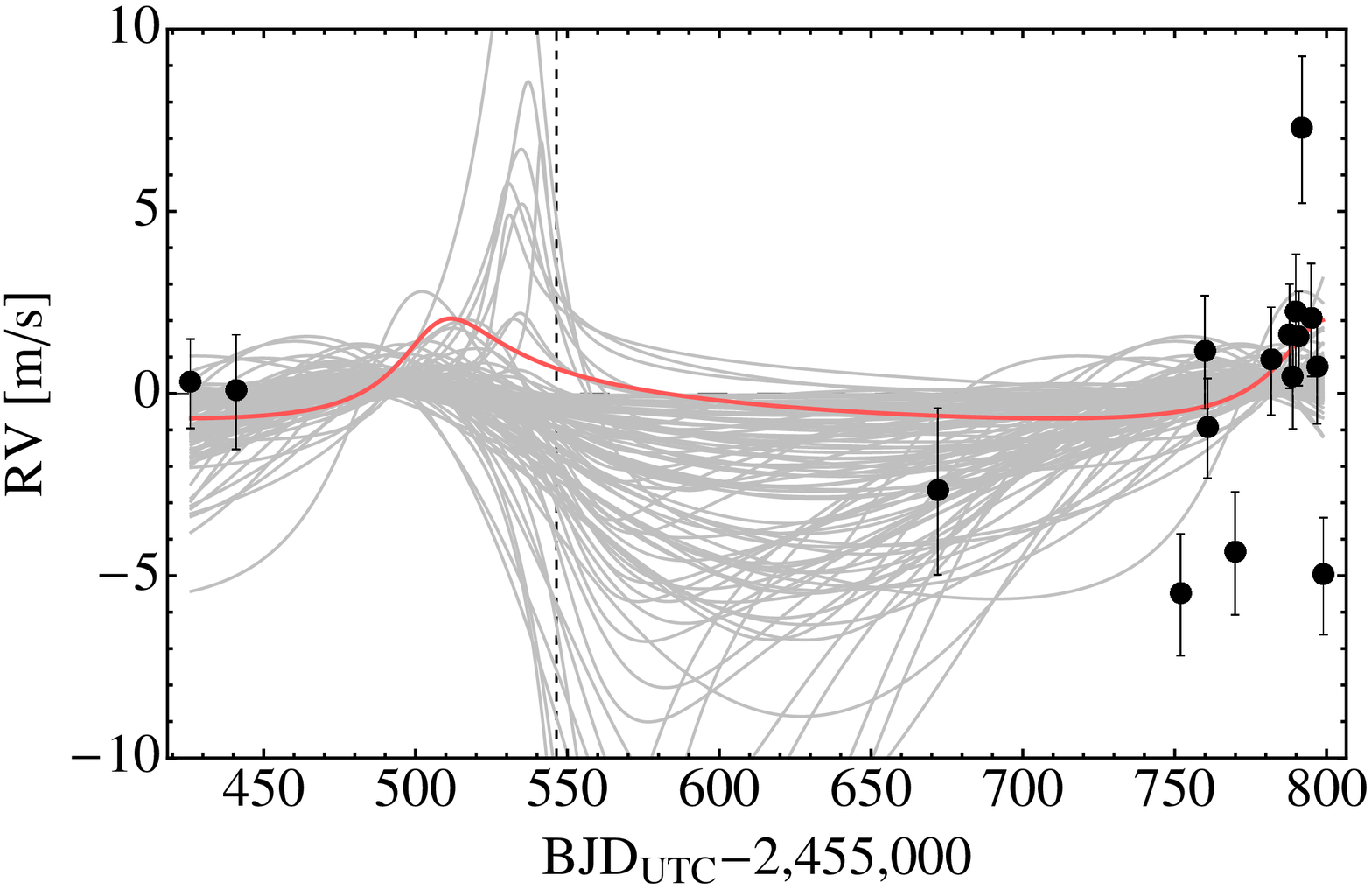}
\caption{\emph{Radial velocities (black points) of \keplerb\ observed by 
B12. We overlay the maximum a-posteriori realization for an 
eccentric orbit planet (model $\mathcal{P}_{\mathrm{LD-free},e_{B*}}$) in red 
and one-hundred random draws from the joint posterior distribution in gray, 
which illustrate the uncertainty in the fit. The dashed vertical line marks the 
time of Q7 transit. The marginalized posterior for $K$, including Bayesian model 
averaging all over planet-only models, yields 
$K=1.6_{-1.1}^{+2.2}$\,m\,s$^{-1}$.
}} 
\label{fig:RVs}
\end{center}
\end{figure}

Our model averaged posteriors also reveal a lower stellar jitter than the
3\,m\,s$^{-1}$ fixed assumption of B12, since we determine
$s = 2.4_{-0.6}^{+0.8}$\,m\,s$^{-1}$. Combining the SAP constrained 
eccentricity, the lower stellar jitter and an overall more refined ephemeris
than B12 (thanks to the extra transit included in this work)
we derive a tighter constraint on the radial velocity semi-amplitude of
$K=1.6_{-1.1}^{+2.2}$\,m\,s$^{-1}$ compared to the B12 value of 
$K=4.9_{-7.4}^{+6.7}$\,m\,s$^{-1}$. Our upper limit on this term, at 95\%
confidence, is $K<6.5$\,m\,s$^{-1}$. Figure~\ref{fig:RVs} shows the maximum
likelihood radial velocity model for a circular orbit model.

\citet{southworth:2007} show how the surface gravity of an exoplanet can be
computed from the transit and RV observables without requiring any assumption
on the stellar/planet physical parameters. Exploiting this trick here, we find
$g_P = 26_{-19}^{+37}$\,m\,s$^{-2}$. Thus, the surface gravity on \keplerb\ is 
most likely greater than that of the Earth but not dramatically so.

Our tighter constraint on $K$ can also be used to derive new constraints on
the planet mass (assuming no moon is present). In order to do so, we require a
stellar mass which we take from the asteroseismology analysis presented in
B12 who found $0.970\pm0.060$\,$M_{\odot}$. Assuming this is
distributed normally, our model averaged posteriors yield 
$M_P = 6.9_{-6.2}^{+20.9}$\,$M_{\oplus}$. For comparison, B12
quote three sets of upper limits for mass at $<36$\,$M_{\oplus}$, 
$<82$\,$M_{\oplus}$ and $<124$\,$M_{\oplus}$ at a confidence of 1\,$\sigma$, 
2\,$\sigma$ and 3\,$\sigma$ respectively. At the same confidence levels, we
estimate $<15$\,$M_{\oplus}$, $<55$\,$M_{\oplus}$ and $<148$\,$M_{\oplus}$
demonstrating the generally much tighter constraint derived here.


\subsection{Constraints on the Radius, Composition \& Atmosphere}

We may also use the stellar radius derived in B12 from
asteroseismology, $R_* = 0.979\pm0.020$, to derive the physical planetary
radius. Our model averaged posteriors yield 
$R_P = 2.396_{-0.181}^{+0.088}$\,$R_{\oplus}$ which is consistent with the
$2.38\pm0.13$\,$R_{\oplus}$ value from B12. Note that two of
the four models over which we have applied the Bayesian model averaging used
free limb darkening coefficients and so are quite robust.

Following the method described in \citet{mah:2013}, one may compute the
minimum atmospheric height ($R_{\mathrm{MAH}}$) for an exoplanet when the
mass and radius are known, by employing a mass-radius relation for a water
dominated planet. Using the mass-radius models of \citet{zeng:2013}, we compute
$R_{\mathrm{MAH}}$ against the 75\%-water-25\%-silicate mass-radius contour
to derive $R_{\mathrm{MAH}} = 0.11_{-0.87}^{+1.04}$\,$R_{\oplus}$ with 54\%
confidence of an atmosphere being present. \keplerb\ is therefore consistent 
with either a pure-water planet with a dense, low atmosphere or a rocky planet 
with a light, extended atmosphere. If \keplerb\ is cloudless, one should expect
significant differences in the transmission spectroscopy signal between these
two hypotheses. Thus, the true composition of \keplerb\ could be determined
from transmission spectroscopy rather than a better mass estimate (although
the latter would clearly work too).


Could the atmosphere of \keplerb\ be observable though? One may estimate the
scale height of the planet's atmosphere using the expression:

\begin{align}
H &= \frac{k_B T_{\mathrm{eq}}}{\mu g_P},
\end{align}

where $k_B$ is the Boltzmann constant, $T_{\mathrm{eq}}$ is the equilibrium 
temperature of the planet and $\mu$ is the mean molecular weight. The planet's 
equilibrium temperature depends upon unknown quantities such as the albedo, 
emissivity and greenhouse effect but ignoring these complexities one may 
estimate $T_{\mathrm{eq}} = 287.7_{-3.3}^{+8.4}$\,K using the simple expression:

\begin{align}
T_{\mathrm{eq}} &= T_{*,\mathrm{eff}} \sqrt{\frac{1}{2 (a_{B*}/R_*) (1-e_{B*}^2)^{1/4}}}.
\end{align}

If \keplerb\ is rocky with a light atmosphere, then we expect $\mu\sim2$ which
leads to $(H/R_P)_{\mathrm{rocky}} = (3.0_{-1.7}^{+7.0})\times10^{-3}$. If 
\keplerb\ is oceanic with a denser atmosphere, then one may adopt an Earth-like 
$\mu$ of $\mu\sim28$ giving 
$(H/R_P)_{\mathrm{oceanic}} = (0.22_{-0.13}^{+0.56})\times10^{-3}$. Translating
these into transit depths, we find a rocky planet has an atmospheric signal
(using Eqn~36 of \citealt{winn:2010}) of 
$\Delta\delta \sim 3.0_{-1.8}^{+7.9}$\,ppm whereas an oceanic world has 
$\Delta\delta \sim 0.2_{-0.1}^{+0.6}$\,ppm. We estimate that it is unlikely that 
these two scenarios could be distinguished with current instrumentation.


\subsection{Constraints on \keplerb's Insolation}

\keplerb\ was originally claimed by B12 to lie within the
habitable-zone of its host star. This statement in fact gave rise to \keplerb's
selection as a high-priority candidate for an exomoon investigation by the
HEK project. Using our revised parameters and a recently revised definition of
the habitable-zone (HZ) from \citet{kopparapu:2013}, we may re-visit this 
important issue. This investigation will purely focus on whether the insolation
received by \keplerb\ places the planet with the habitable-zone, as defined
by \citet{kopparapu:2013}, and will not address other factors affecting
habitability such as the atmospheric composition.

Using the B12 ``best-fit'' parameters, \citet{kopparapu:2013}
estimate \keplerb\ to lie slightly closer to its parent star than the inner-edge 
of the habitable zone, using the moist greenhouse limited model. However, in 
reality the uncertainties on the stellar and planetary parameters can be 
important, especially when an object lies very close to a boundary such as 
\keplerb. We may account for this by drawing a realization from our model 
averaged posteriors and testing whether that particular realization is 
consistent with a habitable planet or not. By simply counting the number of HZ 
realizations, one may phrase the question of habitability in a more 
statistically robust light.

Using the revised habitable-zone (HZ) calculations from \citet{kopparapu:2013}, 
we calculated the number of posterior samples, from the model averaged 
posterior, which could be classified as either ``cold'', ``warm'' or ``hot''. We 
use the maximum-greenhouse case as the outer-edge of the HZ and the 
moist-greenhouse case as the inner-edge. We note that this is the 
``conservative'' habitable-zone since it is based on cloudless models and even 
the Earth falls on the inner-edge boundary for this model. The effective 
temperature and luminosity of the host star were assumed to have normal 
distributions with means and standard deviations given by the values quoted in 
B12. 

\begin{figure}
\begin{center}
\includegraphics[width=8.4 cm]{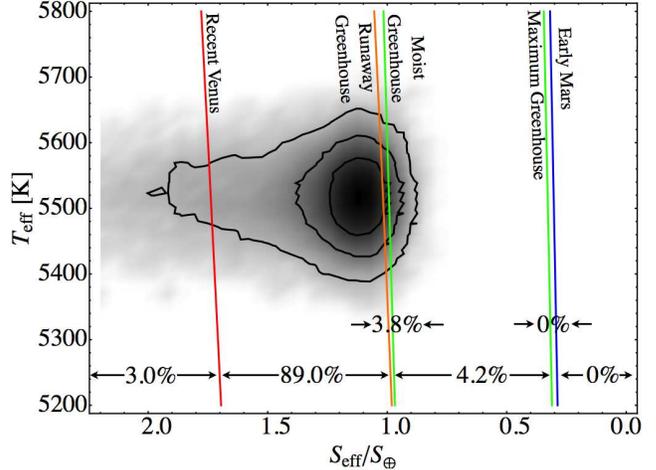}
\caption{\emph{
Joint posterior probability distribution of the stellar effective temperature
($T_{\mathrm{eff}}$) and the insolation received by \keplerb\ relative to that
received by the Earth ($S_{\mathrm{eff}}/S_{\oplus}$). We overlay the various
habitable-zone boundaries derived by \citet{kopparapu:2013} and the fraction of
trials falling between the contours. \keplerb\ is likely warmer than the
Earth but does fall within the empirical habitable-zone boundaries. Posteriors
derived by Bayesian model averaging of all of the planet-only fits.
}} 
\label{fig:HZ}
\end{center}
\end{figure}

We find that 95.8\% of trials were ``hot'' and thus too close to the star to be 
habitable, with the remaining 4.2\% being classed as warm (and thus habitable)
and 0\% cold. The so-called ``empirical'' HZ is defined as the recent-Venus
and early-Mars limits leading to a wider HZ. Using these definitions, we find
that 97.0\% of trials are in the habitable-zone with the remaining 3.0\% of
trials being too hot. In Figure~\ref{fig:HZ} we visualize these constraints
and probabilities to provide a more statistical interpretation of an exoplanet's
habitability.

We therefore conclude that \keplerb\ has a $>95$\% probability of being within 
the empirical HZ but $<5$\% probability of being the conservative HZ, as defined 
by \citet{kopparapu:2013}. We calculate that \keplerb\ receives only 
$13.7_{-8.7}^{+14.6}$\% more insolation than the Earth, as the top of its 
atmosphere. We therefore consider that \keplerb\ can still be considered a
good candidate for being within the habitable-zone, particularly if one
considers models including clouds. We point out that even with an albedo of 
unity, the occultations of \keplerb\ would be undetectable with 
$\delta_{\mathrm{occ}} \simeq 14$\,parts-per-billion.

\begin{table*}
\caption{\emph{
Bayesian evidences of the ten different models (descriptions available in
Table~\ref{tab:models}) attempted to explain the transit light curves of 
\keplerb. $\mathcal{P}$ models are those of a planet-only whereas $\mathcal{S}$ 
models are those of a planet-with-moon. Odds ratios computed relative to the 
strongrest planet-only model. $\sigma$ confidence column is computed relative to 
the best planet-only model ($\mathcal{P}_{\mathrm{LD-free},e_P}$).
}} 
\centering 
\begin{tabular}{l c r c} 
\hline
Model & $\log\mathcal{Z}$ & Odds Ratio & $\sigma$ Confidence \\ [0.5ex] 
\hline
$\mathcal{P}_{\mathrm{LD-prior}}$	& $176828.044 \pm 0.058$ & 0.36 & $-0.91$ \\
$\mathcal{P}_{\mathrm{LD-prior},e_{B*}}$& $176828.797 \pm 0.057$ & 0.77 & $-0.29$ \\
$\mathcal{P}_{\mathrm{LD-free}}$	& $176827.602 \pm 0.059$ & 0.23 & $-1.19$ \\
$\mathcal{P}_{\mathrm{LD-free},e_{B*}}$	& $176829.054 \pm 0.058$ & 1 & - \\
\hline
$\mathcal{S}_{\mathrm{LD-prior}}$	& $176831.704 \pm 0.068$ & 14.15 & 1.81 \\
$\mathcal{S}_{\mathrm{LD-prior},e_{B*}}$& $176832.491 \pm 0.068$ & 31.09 & 2.14\\
$\mathcal{S}_{\mathrm{LD-prior},e_{SB}}$& $176841.248 \pm 0.078$ & 197,599.99 & 4.56 \\
$\mathcal{S}_{\mathrm{LD-free}}$	& $176831.734 \pm 0.069$ & 14.59 & 1.82\\
$\mathcal{S}_{\mathrm{LD-free},e_{B*}}$	& $176832.979 \pm 0.070$ & 50.65 & 2.33 \\
$\mathcal{S}_{\mathrm{LD-free},e_{SB}}$	& $176841.663 \pm 0.078$ & 299,239.64 & 4.65 \\ [1ex]
\hline\hline 
\end{tabular}
\label{tab:evidence} 
\end{table*}

\begin{table*}
\caption{\emph{
Marginalized parameters of \keplerb\ after performing Bayesian
model averaging. We show the results from four different ways of model 
averaging (descriptions available in Table~\ref{tab:models}). Bold-highlighted 
values represent our suggested final results.
Parameters with a ``max'' subscript denote the 95\% upper quantile.
$^{\ddag}$ = Computed using radial velocities and $M_*$ value from B12.
$^{\dagger}$ = Computed using $(M_S/M_P)$ ratio derived purely from moon model 
combined with $M_P$ value derived from radial velocity and $M_*$ value from B12.
$^{\times}$ = Unphysically dense solutions were eliminated in computing 
this term, based upon maximum collisional stripping model from 
\citet{marcus:2010}.
$^{*}$ = Computed using $(M_P/M_*)$ from moon dynamics and $M_*$ value from B12.
$^{\diamond}$ = Computed purely from moon dynamics.
}} 
\centering 
\begin{tabular}{l c c c c} 
\hline
Parameter & $\mathcal{P}_{\mathrm{LD-free},e_{B*}}$ & $\mathbf{<\mathcal{P}_k>}$ & $<\mathcal{P}_k,\mathcal{S}_k>$ for $e_{SB}=0$ & $<\mathcal{P}_k,\mathcal{S}_k>$ \\ [0.5ex] 
\hline
\emph{Planet's transit parameters}\\
\hline
$P_{B*}$\,[days]\dotfill	& $289.86438_{-0.00080}^{+0.00084}$ & $\mathbf{289.86444_{-0.00089}^{+0.0092}}$ & $289.8654_{-0.0013}^{+0.0014}$ & $289.8650_{-0.0011}^{+0.0012}$ \\
$\tau_0$\,[BJD$_{\mathrm{UTC}}-2,455,000$]\dotfill & $546.42471_{-0.00096}^{+0.00102}$ & $\mathbf{546.4248_{-0.0011}^{+0.0011}}$ & $546.4258_{-0.0015}^{+0.0016}$ & $546.4238_{-0.0012}^{+0.0012}$ \\
$(R_P/R_*)$\dotfill	& $0.0219_{-0.0017}^{+0.0010}$ & $\mathbf{0.02254_{-0.00187}^{+0.00054}}$ & $0.02172_{-0.00110}^{+0.00079}$ & $0.02249_{-0.00040}^{+0.00032}$ \\
$(a_{B*}/R_*)$\dotfill	& $186.4_{-1.2}^{+1.2}$ & $\mathbf{186.4_{-1.2}^{+1.2}}$ & $186.4_{-1.2}^{+1.2}$ & $186.4_{-1.1}^{+1.1}$ \\
$b_{B*}$\dotfill	& $0.756_{-0.516}^{+0.081}$ & $\mathbf{0.805_{-0.328}^{+0.020}}$ & $0.786_{-0.156}^{+0.031}$ & $0.8116_{-0.0050}^{+0.0052}$ \\
$i_{B*}$\,[$^{\circ}$]\dotfill	& $89.749_{-0.026}^{+0.095}$ & $\mathbf{89.750_{-0.015}^{+0.016}}$ & $89.7455_{-0.0156}^{+0.0076}$ & $89.7505_{-0.0027}^{+0.0027}$ \\
$\tilde{T}$\,[hours]\dotfill	& $6.953_{-0.089}^{+0.100}$ & $\mathbf{6.972_{-0.080}^{+0.072}}$ & $6.937_{-0.080}^{+0.078}$ & $6.943_{-0.077}^{+0.069}$ \\
$T_{14}$\,[hours]\dotfill	& $7.30_{-0.14}^{+0.14}$ & $\mathbf{7.38_{-0.17}^{+0.10}}$ & $7.30_{-0.13}^{+0.11}$ & $7.391_{-0.074}^{+0.065}$ \\
$T_{23}$\,[hours]\dotfill	& $6.59_{-0.25}^{+0.29}$ & $\mathbf{6.53_{-0.14}^{+0.27}}$ & $6.55_{-0.14}^{+0.17}$ & $6.475_{-0.084}^{+0.074}$ \\
$u_1$\dotfill		& $0.55_{-0.23}^{+0.25}$ & $\mathbf{0.461_{-0.026}^{+0.241}}$ & $0.461_{-0.052}^{+0.241}$ & $0.467_{-0.026}^{+0.343}$ \\
$u_2$\dotfill		& $0.10_{-0.33}^{+0.31}$ & $\mathbf{0.231_{-0.339}^{+0.018}}$ & $0.225_{-0.390}^{+0.019}$ & $0.175_{-0.419}^{+0.065}$ \\
$K$\,[m\,s$^{-1}$]\dotfill	& $1.8_{-1.2}^{+2.7}$ & $\mathbf{1.6_{-1.1}^{+2.2}}$ & $1.6_{-1.1}^{+2.0}$ & $1.39_{-0.95}^{+1.66}$ \\
$s$\,[m\,s$^{-1}$]\dotfill	& $2.36_{-0.59}^{+0.71}$ & $\mathbf{2.41_{-0.63}^{+0.77}}$ & $2.38_{-0.64}^{+0.77}$ & $2.42_{-0.60}^{+0.74}$ \\
$g_P$\,[m\,s$^{-2}$]\,\dotfill	& $29_{-20}^{+45}$ & $\mathbf{26_{-19}^{+37}}$ & $29_{-21}^{+37}$ & $24_{-17}^{+29}$ \\
$e_{B*}\sin\omega_{B*}$\dotfill	& $0.03_{-0.15}^{+0.31}$ & $\mathbf{0.000_{-0.068}^{+0.215}}$ & $0.049_{-0.049}^{+0.204}$ & $0.00_{-0.00}^{+0.00}$ \\
$e_{B*}\cos\omega_{B*}$\dotfill	& $0.08_{-0.27}^{+0.45}$ & $\mathbf{0.00_{-0.12}^{+0.25}}$ & $0.000_{-0.096}^{+0.138}$ & $0.00_{-0.00}^{+0.00}$ \\
$\Psi$\dotfill	& $1.41_{-0.57}^{+2.98}$ & $\mathbf{1.00_{-0.10}^{+2.29}}$ & $1.19_{-0.19}^{+1.15}$ & $1.00_{-0.00}^{+0.00}$ \\
\hline
\emph{Planet's physical parameters}\\
\hline
$R_P$\,[$R_{\oplus}$]\,\dotfill	& $2.32_{-0.17}^{+0.13}$ & $\mathbf{2.396_{-0.181}^{+0.088}}$ & $2.31_{-0.12}^{+0.10}$ & $2.399_{-0.062}^{+0.064}$ \\
$M_P^{\ddag}$\,[$M_{\oplus}$]\,\dotfill	& $7.8_{-6.9}^{+22.0}$ & $\mathbf{6.9_{-6.2}^{+20.9}}$ & $7.2_{-6.4}^{+20.1}$ & $6.6_{-5.9}^{+18.1}$ \\
$M_{P,\mathrm{max}}^{\ddag}$\,[$M_{\oplus}$]\,\dotfill	& $<61.1$ & $\mathbf{<52.8}$ & $<48.9$ & $<43.8$ \\
$\rho_P^{\ddag}$\,[g\,cm$^{-3}$]\,\dotfill	& $2.9_{-2.6}^{+8.8}$ & $\mathbf{2.4_{-2.2}^{+7.5}}$ & $3.2_{-2.9}^{+9.1}$ & $2.6_{-2.3}^{+7.2}$ \\
$\rho_{P,\mathrm{max}}^{\ddag}$\,[g\,cm$^{-3}$]\,\dotfill	& $<25.4$ & $\mathbf{<19.5}$ & $<22.8$ & $<17.4$ \\
$R_{P,\mathrm{MAH}}^{\ddag}$\,[$R_{\oplus}$]\,\dotfill	& $0.00_{-0.88}^{+1.06}$ & $\mathbf{0.11_{-0.87}^{+1.04}}$ & $0.02_{-0.83}^{+1.04}$ & $0.16_{-0.80}^{+1.02}$ \\
$\mathrm{P}(R_{P,\mathrm{MAH}}>0)^{\ddag}$\,\dotfill	& $0.50$ & $\mathbf{0.54}$ & $0.51$ & $0.57$ \\
$S_{\mathrm{eff}}$\,$[S_{\oplus}]$\,\dotfill & $1.19_{-0.12}^{+0.25}$ & $\mathbf{1.137_{-0.087}^{+0.146}}$ & $1.123_{-0.079}^{+0.088}$ & $1.098_{-0.071}^{+0.074}$ \\
$a_{B*}$\,[AU]\,\dotfill	& $0.848_{-0.018}^{+0.018}$ & $\mathbf{0.848_{-0.018}^{+0.018}}$ & $0.848_{-0.018}^{+0.018}$ & $0.848_{-0.018}^{+0.018}$ \\
\hline
\emph{Satellite's transit parameters}\\
\hline
$P_{SB}$\,[days]\dotfill & - & \textbf{-} & $23_{-13}^{+33}$ & $40_{-12}^{+21}$ \\
$\phi_{SB}$\,[$^{\circ}$]\dotfill & - & \textbf{-} & $180_{-110}^{+120}$ & $230_{-210}^{+110}$ \\
$(a_{SP}/R_P)$\dotfill & - & \textbf{-} & $38_{-10}^{+14}$ & $99_{-11}^{+19}$ \\
$\rho_S$\,[g\,cm$^{-3}$]\,\dotfill	& - & \textbf{-} & $1.3_{-1.2}^{+6.4}$ & $0.79_{-0.55}^{+1.24}$ \\
$i_{SB}$\,[$^{\circ}$]\,\dotfill	& - & \textbf{-} & $180_{-120}^{+120}$ & $237_{-155}^{+47}$ \\
$\Omega_{SB}$\,[$^{\circ}$]\,\dotfill	& - & \textbf{-} & $-10_{-130}^{+140}$ & $-35_{-24}^{+170}$ \\
$(R_S/R_P)$\,\dotfill	& $0$ & $\mathbf{0}$ & $0.320_{-0.047}^{+0.047}$ & $0.341_{-0.026}^{+0.027}$ \\
$(M_S/M_P)$\,\dotfill	& $0$ & $\mathbf{0}$ & $0.023_{-0.015}^{+0.022}$ & $0.0029_{-0.0020}^{+0.0030}$ \\
$(M_S/M_P)_{\mathrm{max}}$\,\dotfill	& - & - & $\mathbf{<0.062}$ & $<0.0083$ \\
\hline
\emph{Satellite's physical parameters}\\
\hline
$M_S^{\dagger}$\,[$M_{\oplus}$]\,\dotfill	& $0$ & $\mathbf{0}$ & $0.13_{-0.12}^{+0.58}$ & $0.015_{-0.014}^{+0.064}$ \\
$M_{S,\mathrm{max}}^{\dagger \times}$\,[$M_{\oplus}$]\,\dotfill	& - & - & $\mathbf{<0.54}$ & $<0.16$ \\
$R_S$\,[$R_{\oplus}$]\,\dotfill	& $0$ & $\mathbf{0}$ & $0.74_{-0.10}^{+0.10}$ & $0.819_{-0.062}^{+0.064}$ \\
$R_{S,\mathrm{MAH}}$\,[$R_{\oplus}$]\,\dotfill	& - & \textbf{-} & $-0.06_{-0.49}^{+0.41}$ & $0.41_{-0.27}^{+0.22}$ \\
$\mathrm{P}(R_{S,\mathrm{MAH}}>0)$\,\dotfill	& - & \textbf{-} & $0.44$ & $0.93$ \\
$e_{SB}\sin\omega_{SB}$\dotfill	& - & \textbf{-} & 0 & $0.06_{-0.23}^{+0.18}$ \\
$e_{SB}\cos\omega_{SB}$\dotfill	& - & \textbf{-} & 0 & $0.31_{-0.69}^{+0.22}$ \\
\hline
\emph{Parameters derived from moon model}\\
\hline
$\rho_P^{*}$\,[g\,cm$^{-3}$]\,\dotfill	& - & \textbf{-} & $2.1_{-1.7}^{+8.4}$ & $12.2_{-5.6}^{+8.4}$ \\
$M_P^{*}$\,[$M_{\oplus}$]\,\dotfill	& - & \textbf{-} & $4.6_{-3.8}^{+18.3}$ & $31_{-14}^{+21}$ \\
$M_*^{\diamond}$\,[$M_{\odot}$]\,\dotfill	& - & \textbf{-} & $26_{-26}^{+8635}$ & $0.095_{-0.093}^{+1.711}$ \\
$R_*^{\diamond}$\,[$R_{\odot}$]\,\dotfill	& - & \textbf{-} & $2.9_{-2.5}^{+17.3}$ & $0.43_{-0.30}^{+0.71}$ \\
\hline\hline 
\end{tabular}
\label{tab:params} 
\end{table*}

\section{INJECTED MOON RETRIEVAL}
\label{sec:injected}

\subsection{Generating Synthetic Moon Data}

Before we discuss the results of our search for an exomoon around \keplerb\, we
first demonstrate the sensitivity limits achievable with this data set by
injecting a synthetic moon signal into the data. In preparing this investigation,
we wanted to realistically mimic the exact noise properties of the observed
data i.e. we wish to reproduce all time-correlated noise features. We here
describe several simple steps which enable us to accomplish this.

First, we proceed under the assumption that the real \emph{Kepler} data does not
show evidence for an exomoon (which is in fact our conclusion later in the 
paper). We then take our favored model, $\mathcal{P}_{\mathrm{LD-free}}$, and 
compute the residuals relative to the maximum a-posteriori realization (shown in 
Fig.~\ref{fig:bestfits_es} along with the residuals). We then ``flip'' these 
residuals by simply multiplying the fluxes by $-1$. We then re-add these flipped 
residuals to the maximum a-posteriori realization of model 
$\mathcal{P}_{\mathrm{LD-free}}$ to create a synthetic (but highly realistic) 
planet-only data set, $\mathcal{D}_{P}'$ (where the ``P'' stands for planet-only 
and the dash implies synthetic data). Finally, we repeat this last step but 
instead add the flipped residuals to a pre-determined planet-with-moon model, 
in order to create a synthetic planet-with-moon data set, $\mathcal{D}_{S}'$.

It's important to stress that our technique is only capable of generating a
single realization of synthetic data. However, it precisely reproduces the
exact time-correlated structure of the real data with even the correct phasing
relative to the transit events. Further, we make no assumptions about the
nature of time-correlated noise or its origin, meaning this technique is a
highly robust method for creating a synthetic data set.

The process of fitting for a planet-with-moon photodynamic signal using 
multimodal nested sampling as a computationally demanding one and so we limit 
ourselves to injecting a synthetic moon. For our synthetic moon, we choose a 
mass and radius equal to that of the Earth. The semi-major axis of the moon's 
orbit is chosen such that it resides at the exact same relative distance within 
the Hill sphere that our Moon is separated from the Earth (25.7\% of the Hill
sphere). We choose to use a circular orbit moon and set the
$\cos i_{SB} = \Omega_{SB} = \phi_{SB} = 0$ for simplicity. The planet's
properties are set to that of the maximum a-posteriori values from model
$\mathcal{P}_{\mathrm{LD-free}}$, except for the planet mass where we make
the conservative choice of setting $M_P$ to the 95\% upper quantile of
$45.6$\,$M_{\oplus}$ from the same model (this is conservative since it 
minimizes $M_S/M_P$ and thus any transit timing effects).

\subsection{Fitting the Synthetic Moon Data}

We first start by considering fitting the data $\mathcal{D}_{P}'$, which does
not contain a moon, but does include all of the time-correlated noise features
seen in the real data. In fitting the data, for the sake of computational
expediency, we only attempted two models - that of a circular-orbit planet 
($\mathcal{P}$) and that of a circular-orbit planet with a circular-orbit moon
($\mathcal{S}$). We do not assume coplanarity in the moon fits and allow
the limb darkening coefficients to be fitted for as was done in the real fits.

We find that a planet-only model is strongly favored at 5.0\,$\sigma$ (see
Table~\ref{tab:fakeevidence}), as one should hope for given that no moon was
injected. The 5.0\,$\sigma$ preference also demonstrates the power of Bayesian
model selection - the moon model uses more free parameters and thus is able
to obtain higher likelihoods but the Bayesian evidence naturally penalizes
the model for using these extra free parameters.

\begin{figure*}
\begin{center}
\includegraphics[width=18.0 cm]{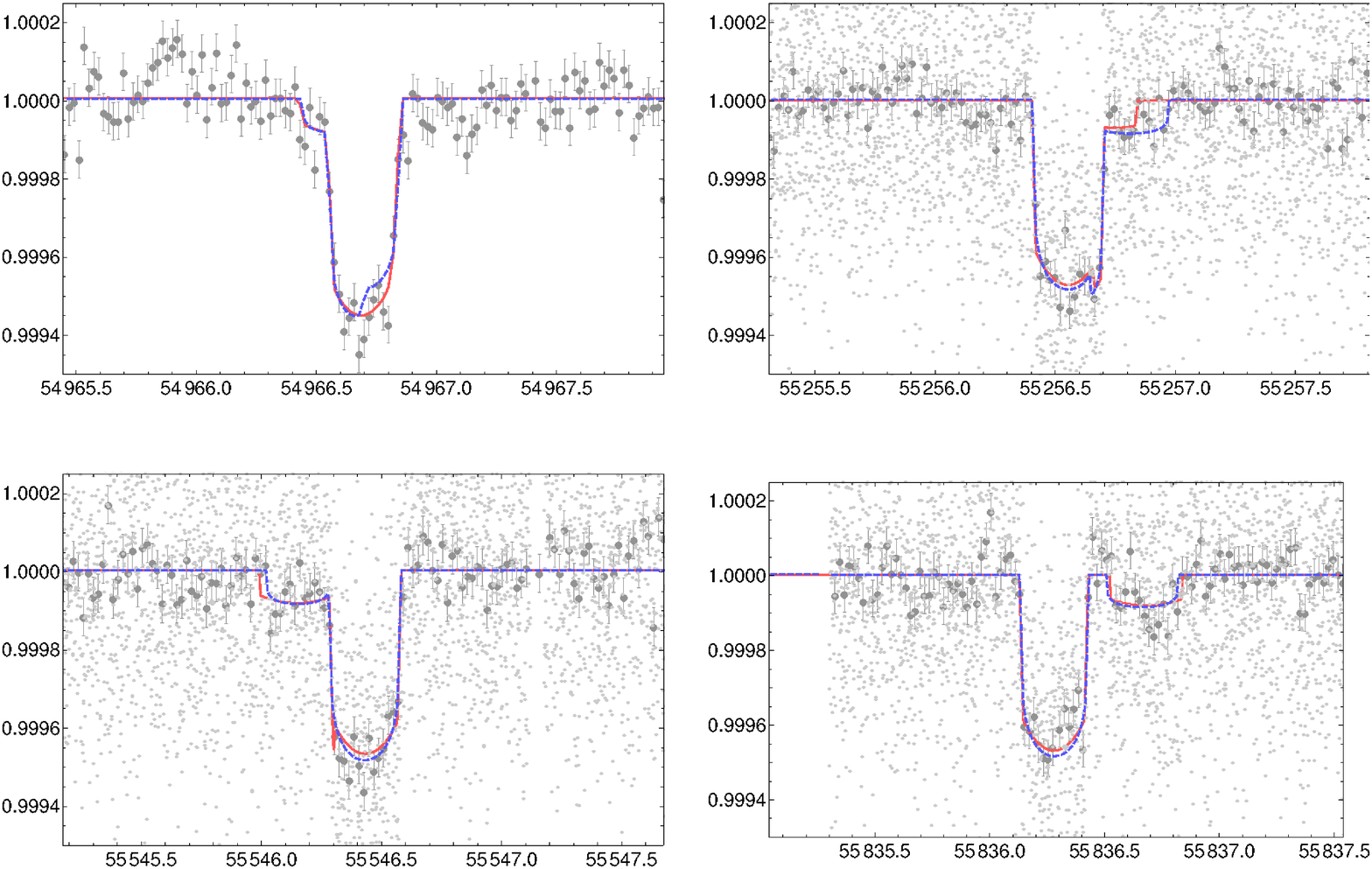}
\caption{\emph{\textbf{Injected moon fits}. From left-to-right then 
top-to-bottom we show the chronological sequence of transits from synthetic data 
set $\mathcal{D}_S'$ for \keplerb. Large dots are the LC-binned data and small 
dots are the SC data. The blue-dashed line shows the ``truth'', which is 
\keplerb\ with an injected Earth-like moon. The red line shows the maximum 
a-posteriori planet-with-moon fit (i.e. a blind retrieval). We are able to 
recover the injected moon to a confidence of 8.3\,$\sigma$. Animations 
of the transits are available at 
https://www.cfa.harvard.edu/$\sim$dkipping/kepler22.html}} 
\label{fig:fakemoon}
\end{center}
\end{figure*}

Focussing on the synthetic planet-with-moon data, $\mathcal{D}_{S}'$, we find
that the planet-with-moon is strongly favored at 8.3\,$\sigma$ (see
Table~\ref{tab:fakeevidence}), again as expected. Whilst we only considered one 
realization for the orbital configuration of the moon, this very significant 
detection demonstrates that an Earth-like moon is readily detectable with the 
current data set. The accuracy of our blind retrieval of this injected moon can 
be visualized by comparing the light curves of the injected truth and the 
maximum a-posteriori blind recovery, as shown\footnote{Animations of the 
transits are also available online at 
https://www.cfa.harvard.edu/$\sim$dkipping/kepler22.html} in 
Figure~\ref{fig:fakemoon}.

\begin{table}
\caption{\emph{
Bayesian evidences two different models attempted to two different synthetic
data sets of \keplerb. $\mathcal{D}_{P}'$ is a synthetic planet-only model
whereas $\mathcal{D}_{S}'$ includes an injected Earth-like satellite.
$\mathcal{P}$ models are those of a planet-only whereas $\mathcal{S}$ models are 
those of a planet-with-moon. Odds ratios and $\sigma$ confidences computed 
relative to the null-hypothesis of a planet-only.
}} 
\centering 
\begin{tabular}{c c c l c} 
\hline
Model & Data & $\log\mathcal{Z}$ & Odds Ratio & $\sigma$ Confidence \\ [0.5ex] 
\hline
$\mathcal{P}$ & $\mathcal{D}_{P}'$ & $176872.912 \pm 0.055$ & 1.00 & - \\
$\mathcal{S}$ & $\mathcal{D}_{P}'$ & $176858.767 \pm 0.073$ & $7.19\times10^{-7}$ & $-4.96$ \\
\hline
$\mathcal{P}$ & $\mathcal{D}_{S}'$ & $176802.104 \pm 0.054$ & 1.00 & - \\
$\mathcal{S}$ & $\mathcal{D}_{S}'$ & $176838.801 \pm 0.077$ & $8.66\times10^{+15}$ & $+8.29$ \\ [1ex]
\hline\hline 
\end{tabular}
\label{tab:fakeevidence} 
\end{table}

\subsection{Habitability of the Injected Moon}

As was done previously for \keplerb\ itself, we should now investigate whether 
the posterior distribution allows us to infer the habitability of our injected 
moon.  Given the complex parameter degeneracies and low-to-moderate 
signal-to-noise of the injected moon signal, it is not immediately clear that we 
are able to definitively determine the moon's habitability, given the broad and 
multimodal nature of the moon parameter posteriors.  

This is further complicated by the fact that moons possess several extra sources 
and sinks of thermal energy compared to their host planets.  If their orbit 
around the planet is elliptical, they can experience tidal heating, which can 
significantly affect surface temperatures even for small eccentricities 
\citep{reynolds:1987,scharf:2006}. They also suffer frequent stellar eclipses by 
their host planet, which can act as an effective energy sink 
\citep{heller:2012}. Both of these phenomena are extremely sensitive to the 
dynamical properties of the planet-moon pair.  As recently demonstrated by 
\citet{forgan:2013} using latitudinal equilibrium balance models (LEBMs), even 
the orbital direction of the moon can have consequences for a telluric exomoon's 
climate.  

The habitable zone of an exomoon is therefore a non-trivial multi-parameter 
manifold, and it is not generally the case that if the host planet is partially 
or completely uninhabitable, the moon must be also.  We therefore cannot rely on 
the habitable zone calculations of \citet{kopparapu:2013} (as was done for 
\keplerb) to determine the habitability of the exomoon. However, this does not 
preclude the possibility of a robust statistical analysis. 

We employ the same LEBMs as described in \citet{forgan:2013} to investigate how 
well the posteriors allow us to recover the moon's habitability. The LEBM allows 
us to evolve the moon's surface temperature as a function of latitude assuming 
it is Earth-like in composition. In the simulation, the moon is subject to 
insolation from the host star, tidal heating from the host planet, eclipses by 
the planet, and infrared atmospheric cooling.  The circulation of heat in the 
atmosphere is modeled by a diffusion equation (see also \citealt{williams:1997} 
and \citealt{spiegel:2008}).

By using each realization of the posteriors as a set of inputs for a LEBM 
simulation, we can construct a distribution of exomoon climates, which can then 
be compared to the climate derived from the ``true'' injected parameters. Each 
LEBM simulation is classified in the following fashion \citep{forgan:2013}:

\begin{enumerate}
\item \emph{\textbf{Habitable Moons}} - these moons possess a habitable surface 
that covers at least 10\% of the total area. This figure is time-averaged, and 
the standard deviation in habitable surface over this time is less than 10\% of 
the mean.
\item \emph{\textbf{Hot Moons}} - these moons display a surface habitability 
fraction of less than 10\%, and typically possess surface temperatures 
above 373\,K across all seasons, and are therefore conventionally uninhabitable.
\item \emph{\textbf{Snowball Moons}} - these moons also display surface 
habitability of less than 10\%, and  have undergone a snowball transition to a 
state where the entire moon is frozen, and are therefore conventionally 
uninhabitable.
\item \emph{\textbf{Variable Moons}} - these moons are similar to the habitable 
moons class above, but the standard deviation in habitable surface over time is 
high, greater than 10\% of the mean.
\end{enumerate}

The simulation run using the injected signal yields a stable, habitable climate 
with a mean temperature of approximately 320\,K. The moon is on a close to 
circular orbit, and hence tidal heating does not play a significant role in the 
resulting climate.  How does this compare to the data retrieved from the 
posteriors?

\begin{figure}
\begin{center}
\includegraphics[width=8.4 cm]{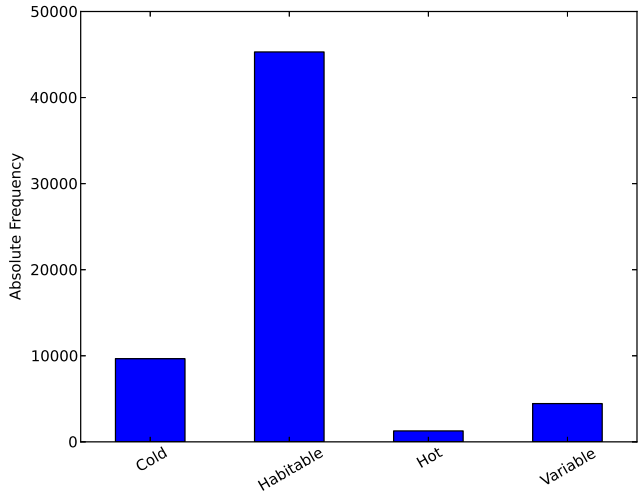}
\caption{\emph{The classification of the LEBM simulations carried out using the 
posterior parameter sets as input. We can see that the majority of the 
simulation runs identify the injected moon as ``habitable'' (i.e. at least 10\% 
of its surface can sustain liquid water, and the standard deviation in this 
value is small). Around one sixth of the simulations are classified as too cold 
to be habitable, with around 2\% of the simulations classifying the moon as too 
hot to be habitable.}} 
\label{fig:habhist_LEBM}
\end{center}
\end{figure}

Figure \ref{fig:habhist_LEBM} shows the resulting classification of all 
posteriors. Around 75\% of all simulations run using the posterior data results 
in a habitable moon. The next most populous classification is the cold moon 
case - this is somewhat unsurprising, as the so-called ``snowball'' transition 
(generated by a rapid increase in albedo as surface water freezes) acts as a 
positive feedback mechanism, making cool climates colder, typically with no 
restorative warming mechanism to combat it. There is no analogous positive 
feedback at the other temperature extreme - indeed, hot climates possess a 
strong negative feedback mechanism through infrared cooling, and as a result the 
hot classification is less populated. As a further consequence of these 
differing feedback mechanisms, there is a small collection of moons which are 
classified as variable, the majority of which tend towards being too hot rather 
than too cold.

\begin{figure}
\begin{center}
\includegraphics[width=8.4 cm]{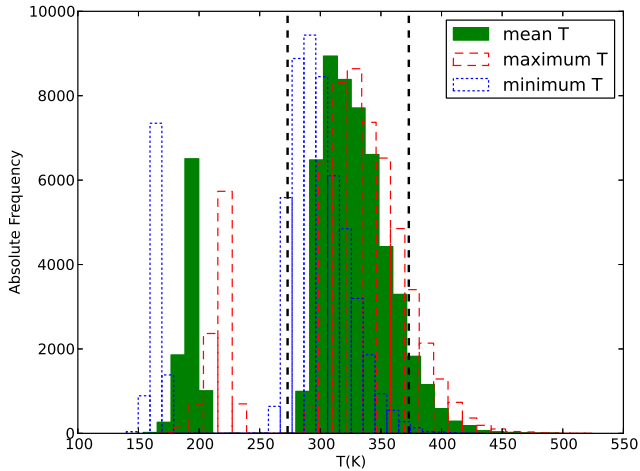}
\caption{\emph{The minimum, maximum and mean temperatures of the LEBM 
simulations carried out using the posterior parameter sets as input. The 
vertical dashed lines indicate the freezing and boiling points of water at 
standard atmospheric pressure, $T=273$\,K and $T=373$\,K respectively.}} 
\label{fig:Tallhist_LEBM}
\end{center}
\end{figure}

We can see this in more detail in Figure \ref{fig:Tallhist_LEBM}, which shows 
the minimum, maximum and mean temperatures exhibited by each simulation. The 
majority of simulations possess minima and maxima which fit inside the 100\,K 
temperature range where water is expected to be liquid on a telluric planet's 
surface.  There is a small tail at $T>373 $ K, but this is not as populous as 
the ``snowball'' population with mean temperatures around 200\,K. The rapid 
albedo transition at 273\,K ensures that few moons can maintain steady surface 
temperatures in this regime, giving rise to the gap centered on $T=250$\,K. The 
mode of the mean temperature distribution is close to the ``true'' mean 
temperature of 320\,K, with a long tail out to around 450\,K.  

In short, we can say with moderate confidence that we can reliably identify 
whether we expect a detected moon to be habitable or otherwise, given the 
posterior distribution, i.e. modeling the surface temperature using the 
posterior data gives similar results to those produced by modeling the injected 
signal. This being said, we should be cognizant of the long tail of the 
resulting temperature distributions produced, and we should also note that one 
in six posterior runs identify the moon as uninhabitable.

\section{PLANET-WITH-MOON RESULTS}

\subsection{Circular Moon Fits}

We first begin by discussing the fits attempted under the assumption that a
putative exomoon follows a circular orbit around the planet \keplerb\ (although
the planet itself is permitted to have an eccentric orbit). In total, there
are four fits we attempted which fall into this category: a circular orbit
planet with and without informative limb darkening priors 
(models $\mathcal{S}_{\mathrm{LD-prior}}$ and $\mathcal{S}_{\mathrm{LD-free}}$
respectively) and an eccentric orbit planet with and without informative
limb darkening priors (models $\mathcal{S}_{\mathrm{LD-prior},e_{B*}}$ and 
$\mathcal{S}_{\mathrm{LD-free},e_{B*}}$ respectively).

None of these four fits show particularly significant model preferences, being
in the range 1-2\,$\sigma$ favorable over a simple planet-only fit (see
Table~\ref{tab:evidence}). As with \citet{hek:2013}, we consider only models 
above $4$\,$\sigma$ to be considered as candidate exomoons. Therefore, we find 
no compelling evidence for a circular orbit exomoon around \keplerb.

\begin{figure*}
\begin{center}
\includegraphics[width=18.0 cm]{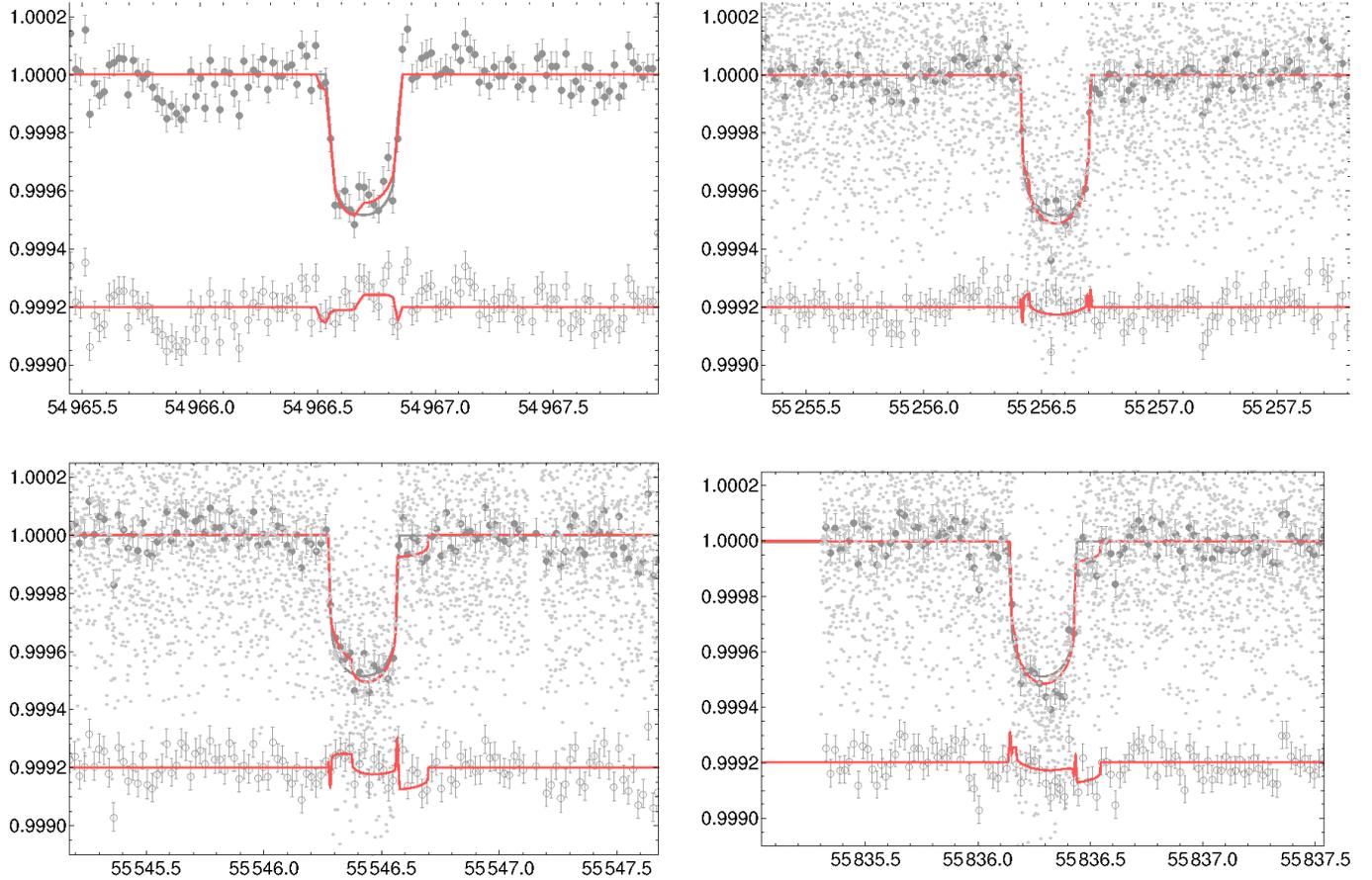}
\caption{\emph{\textbf{Circular moon fits}. From left-to-right then 
top-to-bottom we show the chronological sequence of transits observed by Kepler 
for \keplerb. Large dots are the LC-binned data and small dots are the SC data. 
The gray solid line shows the maximum a-posteriori planet-only fit with the 
corresponding residuals shown offset below. The red line shows the maximum 
a-posteriori planet-with-moon fit from model 
$\mathcal{S}_{\mathrm{LD-free},e_{B*}}$. In the residuals, we also show the 
difference between the two fitted models in red for comparison. Note that this 
line is not equal to the flux change which would be caused by the moon component 
in isolation, rather it is purely the difference between the maximum 
a-posteriori planet-only fit and the maximum a-posteriori planet-with-moon 
fit.}} 
\label{fig:bestfits_ep}
\end{center}
\end{figure*}

Out of the four models, $\mathcal{S}_{\mathrm{LD-free},e_{B*}}$ shows the 
highest significance at 2.3\,$\sigma$ and we show the corresponding maximum 
a-posteriori light curve fit from this model in Figure~\ref{fig:bestfits_ep},
which has a $\Delta\chi^2=41.5$ better fit for $N=25971$ data points.
Although the signal is not significant, we note that the solution corresponds to
a moderately eccentric planet of $e_{B*} = 0.29_{-0.15}^{+0.12}$ and mass
$M_P = 6.5_{-5.5}^{+20.4}$\,$M_{\oplus}$ (derived purely from moon dynamical
model, not from RVs), which is consistent with that derived from the RVs.
The corresponding satellite would have a radius of
$R_S = 0.773_{-0.090}^{+0.083}$\,$R_{\oplus}$ and mass of
$M_S = 0.23_{-0.20}^{+0.80}$\,$M_{\oplus}$ with a broad density posterior
spanning $\rho_S = 2.9_{-2.5}^{+9.5}$\,g\,cm$^{-3}$. We estimate that only
64.7\% of trials yield a physically plausible density by using the internal
structure models of \citet{zeng:2013}.

As with our previous survey \citep{hek:2013}, despite finding a non-detection
we derive upper limits on the mass-ratio for a putative exomoon. Applying
Bayesian model averaging over the four planet-only models and the four
circular-orbit moon models (i.e. eight models in total) allows us to compute
a model-averaged posterior for the mass ratio of 
$(M_S/M_P) = 0.023_{-0.015}^{+0.022}$. The 95\% upper quantile of this posterior
gives $(M_S/M_P)<0.062$ i.e. 6.2\%. For the radius-ratio posterior, $(R_S/R_P)$, 
we note that 97.8\% of trials were positive (recall we explored negative radius 
solutions), corresponding to 2.0\,$\sigma$ significance using the odds-ratio
test given by Equation~4 in \citet{hatp24:2010}.

Unlike previous planets studied, we have a reliable mass constraint for the
host planet from the radial velocities meaning we can translate this mass
ratio into a physical mass upper limit. Further more, \keplerb\ is a validated
small-radius planet and so must conform with physically plausible mass-radius
constraints. We therefore generate a posterior for $M_S$ in absolute units
and then eliminate any trials where either the joint posterior of $M_P$-$R_P$
or $M_S$-$R_P$ corresponds to a density exceeding the maximum mass stripping
limit of an iron-rich planet derived by \citet{marcus:2010}. We find that this
eliminates ~5\% of the posterior trials. From this, we estimate that 
$M_S < 0.54$\,$M_{\oplus}$ to 95\% confidence. We therefore conclude that
\keplerb\ is highly unlikely to host an Earth-like moon on a circular orbit.

\subsection{Eccentric Moon Fits}

Both of the eccentric moon models, $\mathcal{S}_{\mathrm{LD-prior},e_{SB}}$ and
$\mathcal{S}_{\mathrm{LD-free},e_{SB}}$, are strongly favored over the 
planet-only models at $>4$\,$\sigma$ and thus are significant enough to be 
further considered as a candidate, with the latter model being slightly 
preferred. In a $\chi^2$ sense, the maximum a-posteriori realization from model 
$\mathcal{S}_{\mathrm{LD-free},e_{SB}}$ is superior to planet-only fit 
$\mathcal{P}_{\mathrm{LD-free},e_{B*}}$ at $\Delta\chi^2=79.8$ and the circular 
moon fit $\mathcal{S}_{\mathrm{LD-free}}$ at $\Delta\chi^2=38.3$.

\begin{figure*}
\begin{center}
\includegraphics[width=18.0 cm]{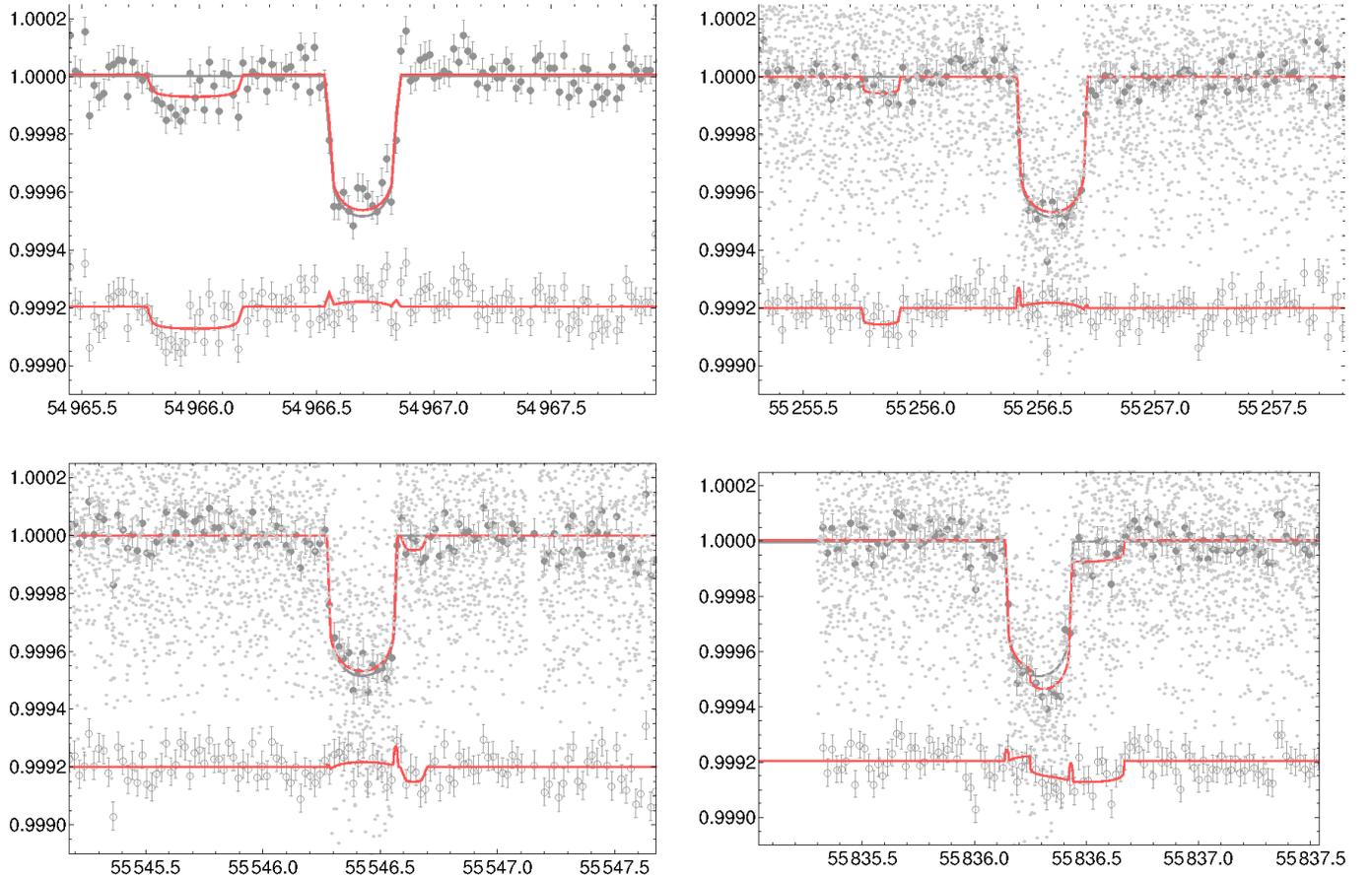}
\caption{\emph{\textbf{Eccentric moon fits}. From left-to-right then 
top-to-bottom we show the chronological sequence of transits observed by Kepler 
for \keplerb. Large dots are the LC-binned data and small dots are the SC data. 
The gray solid line shows the maximum a-posteriori planet-only fit with the 
corresponding residuals shown offset below. The red line shows the maximum 
a-posteriori planet-with-moon fit from model 
$\mathcal{S}_{\mathrm{LD-free},e_{SB}}$. In the residuals, we also show the 
difference between the two fitted models in red for comparison. Note that this 
line is not equal to the flux change which would be caused by the moon component 
in isolation, rather it is purely the difference between the maximum 
a-posteriori planet-only fit and the maximum a-posteriori planet-with-moon 
fit.}} 
\label{fig:bestfits_es}
\end{center}
\end{figure*}

The light curve fits appear to be heavily influenced by a transit-like feature 
occurring prior to the transit of \keplerb\ in the long-cadence data of Q1 (see 
Fig~\ref{fig:bestfits_es}). Unlike the circular orbit fits, the eccentric fits 
have the flexibility to both explain this feature and yet remain comptable
with the rest of the time series. Although our Bayesian model selection approach 
penalizes models for using more free parameters, such as these eccentric fits, 
our likelihood function assumes pure Gaussian noise and so we must remain 
cautious in interpreting what could simply be a time-correlated noise feature of 
unknown origin. This is particularly salient in light of the fact \cofiam\ 
identified Q1 has having the greatest degree of autocorrelation out of the four 
transits used in this work (see \S\ref{sub:cofiam}).

One of the key tests described in \citet{hek:2013} for vetting such systems is
to inspect the derived posteriors and see if they are physically plausible or
not. The Bayesian model averaged posteriors of all ten models attempted are
dominated by the eccentric moon fits due to the large odds ratios and so we
inspect these posteriors to vet these solutions. The dynamically derived
planet density appears rather high at $\rho_P =12.2_{-5.6}^{+8.4}$\,g\,cm$^{-3}$
yielding $M_P = 31_{-14}^{+21}$\,$M_{\oplus}$. This may be compared with the RV 
derived $M_P$ from the same posteriors of 
$M_P = 6.5_{-5.8}^{+17.2}$\,$M_{\oplus}$ and $M_P<40.4$ to 95\% confidence, from
which we conclude the result is slightly incompatible.

Before commenting on the putative exomoon's composition, we note that the orbit
appears excited with $e_{SB} = 0.46_{-0.22}^{+0.12}$ and inclined 
$\sim15^{\circ}$ (highly multimodal posterior) from the planet's orbital plane. 
For the composition, we find an unusually low density of 
$\rho_S = 0.79_{-0.55}^{+1.24}$\,g\,cm$^{-3}$, which favors a bulk density below 
that of water/ice. Using the minimum atmospheric height method of 
\citet{mah:2013} and using the derived moon radius of 
$R_S = 0.818_{-0.062}^{+0.065}$\,$R_{\oplus}$, we find that 93.3\% of trials 
yield a radius exceeding a pure water/ice composition moon. The only way to
explain this situation is an extended atmosphere but given that
$M_S = 0.015_{-0.013}^{+0.062}$\,$M_{\oplus}$, the moon would rapidly lose an 
atmosphere, given the equilibrium temperature of $286\pm3$\,K. 
Given that \kepler\ is not a young star (B12) and so presumably 
the exomoon has had plenty of time for an atmosphere to have escaped already, we 
must invoke a continuously replenishing atmosphere to explain these values. 

\begin{figure*}
\begin{center}
\includegraphics[width=18.0 cm]{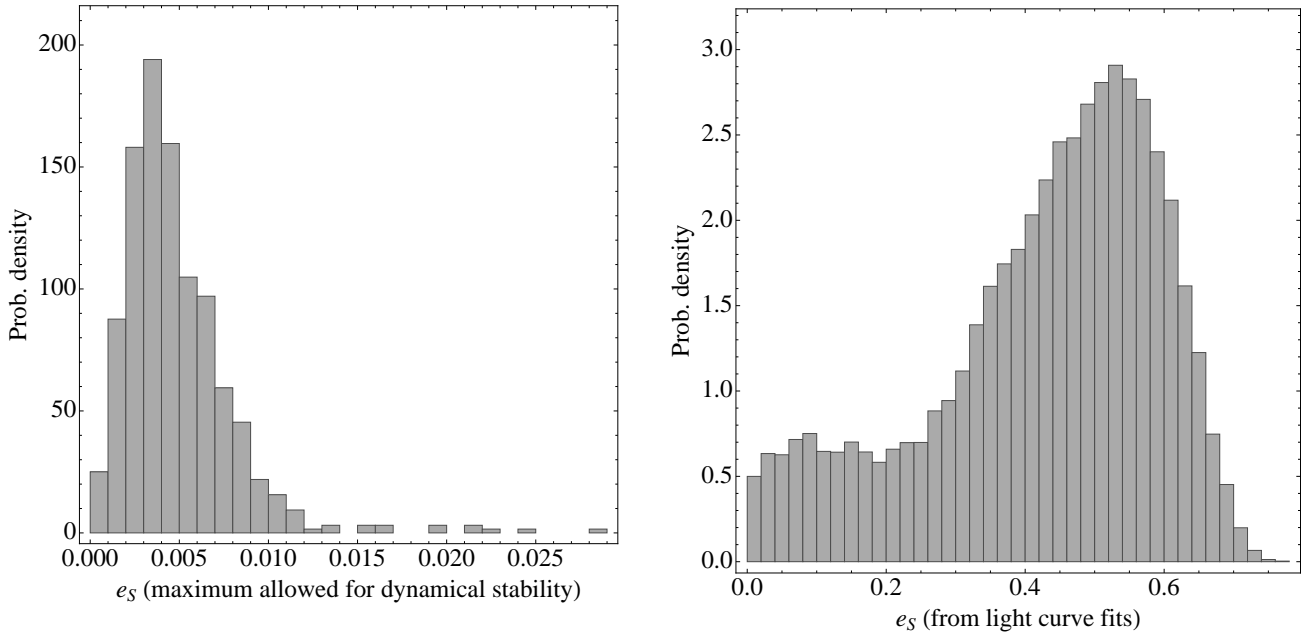}
\caption{\emph{
\textbf{Left:} Using Equation~25 of \citet{donnison:2010}, we calculate the 
posterior distribution of the maximum allowed exomoon eccentricity for Hill 
stability. \textbf{Right:} Derived distribution of the light curve fitted 
exomoon eccentricity from the model averaged posteriors, for which 99.97\% of 
trials are Hill unstable.
}} 
\label{fig:eshisto}
\end{center}
\end{figure*}

The orbit of the inclined, eccentric moon can also be shown to be unstable. 
The Hill stability of inclined low-mass binaries in the three body problem was
explored extensively in \citet{donnison:2010}. \citet{donnison:2010} presented
expressions for the maximum stable eccentricity of an inclined binary in the 
case where the combined planet-moon mass is much less than that of the host 
star, as is applicable for the \keplerb\ system. Using Equation~25 of this work,
which has a complex dependency with $M_S$, $M_P$, $M_*$, $i_{SB}$, $a_{B*}$ and 
$a_{SP}$, we estimate that 99.97\% of the joint posterior realizations exhibit
Hill instability (see Figure~\ref{fig:eshisto}). Driving this determination
is the wide-separation of the putative moon's orbit of 
$a_{SP} \simeq 100$\,$R_P$ combined with the inclined, eccentric nature of the
orbit.

We therefore consider that the solution is a false-positive, likely induced by 
residual autocorrelation in the long-cadence data of Q1. This false transit
signal can only be fitted by driving to high eccentricities. One effect of high
exomoon eccentricity is to increase the r.m.s. TTV and TDV amplitudes induced
by the moon on the planet \citep{kipping:2009a}. Since no TTVs or TDVs occur in 
the \keplerb\ data (Figure~\ref{fig:TTVs}), the algorithm is forced to set the 
moon to be a very low-mass object in order to explain the full data set.

\subsection{Excluded Moon Mass}

It has been shown that the $>4$\,$\sigma$ eccentric exomoon signal is a false
positive i.e. spurious detection. Spurious detections cannot be used to derive
upper limits for reasons discussed in \citet{hek:2013} and so we do not use
these results to derive our excluded moon mass limits presented here. This is
equivalent to assigning a negligible prior model probability to models
$\mathcal{S}_{\mathrm{LD-free},e_{SB}}$ and 
$\mathcal{S}_{\mathrm{LD-prior},e_{SB}}$. Performing Bayesian model averaging 
over the remaining eight models provides reliable upper limits on the presence 
of an exomoon. We find that $(M_S/M_P)<6.2$\% to 95\% confidence which can be 
converted into a physical mass estimate by leveraging the $(M_P/M_*)$ constraint 
from the radial velocities and $M_*$ from asteroseismology to give 
$M_S<0.54$\,$M_{\oplus}$ to 95\% confidence.

\section{SUMMARY}
\label{sec:summary}

Due to the large number of results in this paper, we summarize the most
important findings below:

\begin{itemize}
\item[{\tiny$\blacksquare$}] We have conducted the first search for an exomoon
around a habitable-zone exoplanet and find no compelling evidence for a 
companion to \keplerb. Furthermore, we constrain $M_S<0.54$\,$M_{\oplus}$ to 
95\% confidence.
\item[{\tiny$\blacksquare$}] We have demonstrated that an Earth-like moon
would be detectable around \keplerb\ with the current data to 8.3\,$\sigma$
by signal injection and recovery.
\item[{\tiny$\blacksquare$}] We have further shown that such an injected moon
would very likely have conditions suitable for liquid water on the surface
through latitudinal energy balance modelling (LEBM) of the posteriors samplings.
\item[{\tiny$\blacksquare$}] We have introduced several new improvements to the
HEK methodology, including exploring negative-radius, retrograde and eccentric
moon solutions, free limb darkening sampled from a Dirichlet prior and Bayesian 
model averaging.
\item[{\tiny$\blacksquare$}] For the first time, the radial velocities of
\keplerb\ have been fitted with a model accounting for free eccentricity and
free stellar jitter. Combined with an updated ephemeris from a new transit in
Q11, we derive a tighter mass constraint on \keplerb\ of 
$M_P<52.8$\,$M_{\oplus}$ to 95\% confidence.
\item[{\tiny$\blacksquare$}] Utilizing Single-body Asterodensity Profiling (SAP)
and the radial velocities, we provide the first constraints on the orbital
eccentricity of \keplerb\ with $e_{B*}=0.13_{-0.13}^{+0.36}$ and $e_{B*}<0.71$ 
to 95\% confidence.
\item[{\tiny$\blacksquare$}] From our refined transit model, we estimate that
\keplerb\ has a $>95$\% probability of lying within the empirical habitable-zone
but a $<5$\% probability of lying within the conservative habitable-zone (as
defined by \citealt{kopparapu:2013}). We derive an insolation of 
$S_{\mathrm{eff}} = 1.137_{-0.087}^{+0.146}$\,$S_{\oplus}$.
\end{itemize}

The above results were derived through photodynamical modeling and multimodal
nested sampling regression requiring 49.7\,years of CPU time$^{9}$. Compared to 
other systems analyzed during the HEK project \citep{nesvorny:2012,hek:2013}, 
this load is higher than average due to the high-resolution fitting mode 
employed here (see \S\ref{sub:hires}). Nevertheless, this highlights the unique 
computational challenges of seeking exomoons.

We find no evidence for an Earth-like exomoon around \keplerb\ and yet have 
shown that the present data can easily detect such an object via signal 
injection. Current observations therefore dictate that \keplerb\ does not 
possess an Earth-like habitable moon. Our results then, combined with the
very robust measurement of the planet's radius, mean that \kepler\ does not
possess an Earth analog. This does not mean that the system possesses no
options for an inhabited world, with notable possiblities being a smaller,
presently undetectable moon (e.g. $M_S\sim0.2$\,$M_{\oplus}$) or a possible
ocean on \keplerb. However, it is now clear that this is not the location to 
find a second Earth.

To date, nine systems have been surveyed for exomoons by the HEK project
\citep{nesvorny:2012,hek:2013} and no detections have been made with most
cases yielding detection sensitivities of $\sim M_{\oplus}$. We caution that the
number of systems analyzed remains too small to draw any meaningful conclusions
about the occurrence of large moons, $\eta_{\leftmoon}$, but this is the 
ultimate goal of our project. In coming work, two future surveys will focus on
i) M-dwarf host star planetary candidates and ii) planets exhibiting repeated
visual anomalies. Slowly then, the landscape of the frequency of exomoons will
be revealed.

\acknowledgements
\section*{Acknowledgements}

This work made use of the Michael Dodds Computing Facility.
We thank the anonymous reviewer for their thoughtful comments which improved
the quality of our manuscript.
DMK is funded by the NASA Carl Sagan Fellowships. DF gratefully acknowledges 
support from STFC grant ST/J001422/1. JH and GB acknowledge partial 
support from NSF grant AST-1108686 and NASA grant NNX12AH91H. DN acknowledges 
support from NSF AST-1008890.



\end{document}